\documentclass[sigplan,nonacm,pbalance]{acmart}

\AtBeginDocument{%
  }

\setcopyright{cc}
\setcctype{by}
\copyrightyear{2026}

\usepackage{holtexbasic}
\renewcommand{\HOLTokenTurnstile}{\ensuremath{\vdash\rule{-0.19em}{0em}}}
\renewcommand{\HOLConst}[1]{\textsf{#1}}
\renewcommand{\HOLFieldName}[1]{\textsf{#1}}
\renewcommand{\HOLSymConst}[1]{\HOLConst{#1}}
\renewcommand{\HOLTyOp}[1]{\HOLConst{#1}}
\renewcommand{\HOLinline}[1]{\textsf{\ensuremath{#1}}}
\renewcommand{\HOLKeyword}[1]{\mathsf{#1}}
\renewcommand{\HOLTokenBar}{\ensuremath{\mathtt{|}}}

\renewcommand{\HOLTokenLeftrec}{\ensuremath{\langle\!|\;}}
\renewcommand{\HOLTokenRightrec}{\ensuremath{\;|\!\rangle}}

\newcommand{\graycomment}[1]{\textcolor{black!65}{#1}}

\usepackage{environ}
\NewEnviron{holthmenv}{%
  \[
  \scalebox{1.0}{\begin{array}[t]{l}
  \BODY
  \end{array}}
  \]}

\usepackage{alltt}

\usepackage{proof}
\usepackage{placeins}

\begin{document}

\title{Verified VCG and Verified Compiler for Dafny}

\author{Daniel Nezamabadi}
\orcid{0009-0006-8590-435X}
\affiliation{%
  \institution{ETH Zurich}
  \country{Switzerland}
}

\author{Magnus O. Myreen}
\orcid{0000-0002-9504-4107}
\affiliation{%
  \institution{Chalmers University of Technology\\University of Gothenburg}
  \country{Sweden}
}

\author{Yong Kiam Tan}
\orcid{0000-0001-7033-2463}
\affiliation{%
  \institution{Institute for Infocomm Research (I$^2$R), A*STAR}
  \country{Singapore}
}
\affiliation{%
  \institution{Nanyang Technological University}
  \country{Singapore}
}

\begin{abstract}
  Dafny is a verification-aware programming language that comes with a compiler and static program verifier.
  However, neither the compiler nor the verifier is proved correct; in fact, soundness bugs have been found in both tools.
  This paper shows that the aforementioned Dafny tools can be developed with foundational correctness guarantees.
  We present a functional big-step semantics for an imperative subset of Dafny and, based on this semantics, a verified verification condition generator (VCG) and a verified compiler for Dafny.
  The subset of Dafny we have formalized includes mutually recursive method calls, while loops, and arrays---these language features are significant enough to cover challenging examples such as McCarthy's 91 function and array-based programs that are used when teaching Dafny.
  The verified VCG allows one to prove functional correctness of annotated Dafny programs,
  while the verified compiler can be used to compile verified Dafny programs to CakeML programs.
  From there, one can obtain executable machine code via the (already verified) CakeML compiler, all while provably maintaining the functional correctness guarantees that were proved for the source-level Dafny programs.
  Our work has been mechanized in the HOL4 theorem prover.
\end{abstract}



\maketitle

\section{Introduction}
Dafny~\cite{dafny} is a verification-aware programming language with built-in support for specifications and supporting tools: a compiler and static program verifier.
Recently, Dafny has been successfully used by Amazon Web Services to develop a new and improved version of their authorization engine while provably maintaining the functional behavior of the original implementation~\cite{dafny-authorization-engine}.

To build an intuition on how Dafny works, consider McCarthy's 91 function~\cite{91-function} shown in Figure~\ref{code:dfy:mccarthy}, a nested recursive function that is used as a basic test for automated verification.
It is interesting from an automation perspective because to show that the value of the \verb|decreases| clause actually decreases in the second recursive call, the prover must simultaneously establish the postcondition for \verb|tmp|.
To prove that the implementation satisfies the specifications given in the \verb|ensures| and \verb|decreases| clauses, the verifier translates Dafny to Boogie~\cite{boogie, boogie2}, an intermediate verification language and verification condition generator (VCG).
The generated verification conditions are then checked using a satisfiability modulo theories (SMT) solver, such as Z3~\cite{z3}.
In order to run the program, the Dafny compiler translates it to a language, e.g., C\#, whose toolchain(s) can then be used to generate a binary.

\begin{figure}
\centering
\begin{alltt}
method M(n: int) returns (r: int)
  ensures r == if n <= 100 then 91 else n - 10
  decreases 111 - n
\{
  if n <= 100 \{
    var tmp := M(n + 11);
    r := M(tmp);
  \} else \{
    r := n - 10;
  \}
\}
\end{alltt}
\caption{``91 function'' in Dafny.\protect\footnotemark}
\label{code:dfy:mccarthy}
\end{figure}
\footnotetext{Adapted from the Dafny GitHub repository.}

Both the Dafny compiler and VCG pipeline are substantial pieces of software that must produce trustworthy results.
Unfortunately, past work has identified issues---particularly soundness bugs---in both the implementation of the compiler and the verifier~\cite{dafny-test, dafny-test-2}.
We aim to address these limitations by showing that it is possible for Dafny to be done in a foundational, end-to-end manner.
In this work, we make the following contributions:
\begin{itemize}
\item Define functional big-step semantics~\cite{functional-big-step} for an imperative subset of Dafny, including mutually recursive method calls, while loops, and arrays (Section~\ref{sec:semantics}).
\item Implement a verified compiler from our subset of Dafny to CakeML~\cite{cakeml}, which, in turn, provides a verified compiler down to machine code (Section~\ref{sec:vcomp}).
\item Define a weakest precondition (wp) calculus for the Dafny subset, prove its soundness, and use it to implement a verified VCG (Section~\ref{sec:vvcg}).
\end{itemize}

Our subset is expressive enough to support examples such as the previously introduced 91 function, and array-based programs that are used when teaching Dafny.
We demonstrate our contributions in Section~\ref{sec:together}, where we apply both the compiler and VCG to the 91 function and a method that swaps two elements in an array (\verb|Swap|),
followed by proving the resulting verification conditions to be valid.
This yields functional correctness guarantees at the level of compiled CakeML functions.

We have mechanized this work in the HOL4 theorem prover.
Below, we present and motivate key design choices in the formalization that enabled each of our contributions.

\section{Semantics} \label{sec:semantics}
Before we can write a verified compiler or verified VCG, we must first formalize the meaning of Dafny programs.
We opted to define an untyped, \emph{functional big-step semantics}~\cite{functional-big-step} for Dafny, i.e., a definitional interpreter function with a clock (also called fuel) to ensure its termination and thereby definability within the HOL4 logic.
This style of semantics is also used for CakeML and its compiler's intermediate languages because it is particularly well-suited for mechanizing compiler correctness results in HOL4~\cite{functional-big-step}.

The main syntactic objects are expressions (\HOLTyOp{dfy_exp}), statements (\HOLTyOp{dfy_stmt}), and programs (\HOLTyOp{program}); Dafny programs are a list of declarations used to look up functions and methods.
Our interpreters for Dafny expressions and statements have the following type signatures, respectively:
\begin{holthmenv}
\HOLConst{evaluate_exp}\;:\;\HOLTyOp{state}\;\ensuremath{\rightarrow}\;\HOLTyOp{dfy_env}\;\ensuremath{\rightarrow}\;\HOLTyOp{dfy_exp}\\
\;\;\;\;\;\;\;\;\;\;\;\;\;\;\;\;\;\;\;\;\;\;\;\;\;\ensuremath{\rightarrow}\;\HOLTyOp{state}\;\HOLTokenProd{}\;\HOLTyOp{value}\;\HOLTyOp{exp_result}\\
\HOLConst{evaluate_stmt}\;:\;\HOLTyOp{state}\;\ensuremath{\rightarrow}\;\HOLTyOp{dfy_env}\;\ensuremath{\rightarrow}\;\HOLTyOp{dfy_stmt}\\
\;\;\;\;\;\;\;\;\;\;\;\;\;\;\;\;\;\;\;\;\;\;\;\;\;\ensuremath{\rightarrow}\;\HOLTyOp{state}\;\HOLTokenProd{}\;\HOLTyOp{stmt_result}
\end{holthmenv}

In this section, we will first introduce the types representing the semantic primitives such as \HOLinline{\HOLTyOp{value}}, \HOLinline{\ensuremath{\alpha}\;\HOLTyOp{exp_result}}, \HOLinline{\HOLTyOp{stmt_result}}, \HOLinline{\HOLTyOp{dfy_env}}, and \HOLinline{\HOLTyOp{state}}.
Afterward, we will define interpreter semantics for Dafny expressions and statements, followed by the semantics of Dafny programs.

\subsection{Semantic Primitives}
\paragraph{Values} \label{semantics:value}
We support integers, booleans, strings, and arrays, and define the value type as follows:
\begin{holthmenv}
  \HOLTyOp{value}\;=\\
\;\;\;\;\HOLConst{IntV}\;\HOLTyOp{int}\\
\;\;\HOLTokenBar{}\;\HOLConst{BoolV}\;\HOLTyOp{bool}\\
\;\;\HOLTokenBar{}\;\HOLConst{StrV}\;\HOLTyOp{string}\\
\;\;\HOLTokenBar{}\;\HOLConst{ArrV}\;\HOLTyOp{num}\;\HOLTyOp{num}\;\HOLTyOp{type}
\end{holthmenv}
Array values hold their length, location on the heap, and the type of their elements.\footnote{While it is not necessary for array values to hold the length of the array in the one-dimensional case, it is necessary for multi-dimensional arrays (for which we plan to add support in the future), as in some cases we cannot determine the length of inner arrays by indexing to them first if they follow zero-length dimensions.\label{fn:arrlen}}
Note that our language supports nested arrays, e.g., an array of arrays of integers, which is not to be confused with multi-dimensional arrays.
The heap itself is a list of heap values; thus, heap locations are indices into that list.
We currently only support arrays but plan to add objects in the future.
\begin{holthmenv}
  \HOLTyOp{heap_value}\;=\;\HOLConst{HArr}\;(\HOLTyOp{value}\;\HOLTyOp{list})\;\HOLTyOp{type}
\end{holthmenv}

\paragraph{Results} \label{semantics:result}
The interpreter can fail due to undefined behavior (\HOLConst{Rfail}), such as trying to read an undeclared or uninitialized variable, or timing out (a feature of functional big-step semantics):
\begin{holthmenv}
  \HOLTyOp{err_result}\;=\;\HOLConst{Rfail}\;\HOLTokenBar{}\;\HOLConst{Rtimeout}
\end{holthmenv}

The result type is different for expressions and statements.
Evaluating an expression can either return a value or fail due to the reasons above:
\begin{holthmenv}
  \ensuremath{\alpha}\;\HOLTyOp{exp_result}\;=\;\HOLConst{Rval}\;\ensuremath{\alpha}\;\HOLTokenBar{}\;\HOLConst{Rerr}\;\HOLTyOp{err_result}
\end{holthmenv}
By parameterizing the type of the value being returned, we can use the same data type when returning a single value or a list of values, with the latter occurring when we evaluate a list of expressions.

When evaluating a statement, the result can be to either continue or stop the current evaluation:
\begin{holthmenv}
  \HOLTyOp{stmt_result}\;=\;\HOLConst{Rcont}\;\HOLTokenBar{}\;\HOLConst{Rstop}\;\HOLTyOp{stop}
\end{holthmenv}
The current evaluation can be stopped due to evaluating a return statement or an error:
\begin{holthmenv}
  \HOLTyOp{stop}\;=\;\HOLConst{Sret}\;\HOLTokenBar{}\;\HOLConst{Serr}\;\HOLTyOp{err_result}
\end{holthmenv}
By factoring out the reason for stopping into a separate type, we make it easier in the future to add support for more statements that cause changes in the control flow, such as \verb|continue| and labeled break statements.

\paragraph{Environments} \label{semantics:environment}
The semantic environment record contains the entire Dafny program, which is searched when calling functions or methods.
\begin{holthmenv}
  \HOLTyOp{dfy_env} = \HOLTokenLeftrec{}\;\HOLFieldName{prog}\;:\;\HOLTyOp{program}\;\HOLTokenRightrec{}
\end{holthmenv}

\paragraph{States} \label{semantics:state}
We define the state of the interpreters with the following record:
\begin{holthmenv}
  \HOLTyOp{state} = \HOLTokenLeftrec{}\\
\;\;\;\HOLFieldName{clock}\;:\;\HOLTyOp{num};\\
\;\;\;\HOLFieldName{locals}\;:\;(\HOLTyOp{string}\;\HOLTokenProd{}\;\HOLTyOp{value}\;\HOLTyOp{option})\;\HOLTyOp{list};\\
\;\;\;\HOLFieldName{heap}\;:\;\HOLTyOp{heap_value}\;\HOLTyOp{list};\\
\;\;\;\HOLFieldName{locals_old}\;:\;(\HOLTyOp{string}\;\HOLTokenProd{}\;\HOLTyOp{value}\;\HOLTyOp{option})\;\HOLTyOp{list};\\
\;\;\;\HOLFieldName{heap_old}\;:\;\HOLTyOp{heap_value}\;\HOLTyOp{list};\\
\;\;\;\HOLFieldName{locals_prev}\;:\;(\HOLTyOp{string}\;\HOLTokenProd{}\;\HOLTyOp{value}\;\HOLTyOp{option})\;\HOLTyOp{list};\\
\;\;\;\HOLFieldName{heap_prev}\;:\;\HOLTyOp{heap_value}\;\HOLTyOp{list}\\
\HOLTokenRightrec{}
\end{holthmenv}

To ensure that the functional big-step interpreters are well-defined, terminating functions, the \HOLConst{clock} keeps track of the number of clock ticks that are still available.
Each time we evaluate an expression or statement that could potentially diverge, such as calls or while loops, we first check whether there are ticks still available.
If no more ticks are available, the interpreter returns the timeout error as its result.
Otherwise, the clock is decremented, and evaluation continues.

The \HOLConst{locals} field is an association list that keeps track of all declared variables and their values.
By using the first match when looking up a variable, we can support shadowing by simply prepending elements.
Since Dafny does not require local variables to be initialized immediately when they are declared, we use an option to indicate their initialization status.
Recall that trying to read an undeclared or uninitialized variable causes the interpreter to return the \HOLConst{Rfail} error.
As mentioned earlier, the \HOLConst{heap} field models a list of heap values, where elements are looked up via their (list) index.

The \HOLConst{locals_old} and \HOLConst{heap_old} fields hold the values of \HOLConst{locals} and \HOLConst{heap} right before executing the body of a method.
These are necessary to support the semantics of Dafny's \verb|old|-expressions, which are discussed in the next section.

Finally, the \HOLConst{locals_prev} and \HOLConst{heap_prev} fields are used internally by the generated verification conditions, allowing them to refer to a previously set state as needed.
\subsection{Expressions and \texttt{old}-expressions} \label{semantics:old}
The semantics of Dafny expressions is straightforward.
For example, we define the semantics for a unary operation as follows by first evaluating the operand to a value, then applying the unary operator:
\begin{holthmenv}
  \HOLConst{evaluate_exp}\;\HOLFreeVar{st\ensuremath{\sb{0}}}\;\HOLFreeVar{env}\;(\HOLConst{UnOp}\;\HOLFreeVar{uop}\;\HOLFreeVar{e})\;\HOLTokenDefEquality{}\\
\;\;\HOLKeyword{case}\;\HOLConst{evaluate_exp}\;\HOLFreeVar{st\ensuremath{\sb{0}}}\;\HOLFreeVar{env}\;\HOLFreeVar{e}\;\HOLKeyword{of}\\
\;\;\;\;(\HOLBoundVar{st\ensuremath{\sb{1}}}\HOLSymConst{,}\HOLConst{Rval}\;\HOLBoundVar{v})\;\HOLTokenImp{}\\
\;\;\;\;\;\;(\HOLKeyword{case}\;\HOLConst{do_uop}\;\HOLFreeVar{uop}\;\HOLBoundVar{v}\;\HOLKeyword{of}\\
\;\;\;\;\;\;\;\;\;\HOLConst{None}\;\HOLTokenImp{}\;(\HOLBoundVar{st\ensuremath{\sb{1}}}\HOLSymConst{,}\HOLConst{Rerr}\;\HOLConst{Rfail})\\
\;\;\;\;\;\;\;\HOLTokenBar{}\;\HOLConst{Some}\;\HOLBoundVar{res}\;\HOLTokenImp{}\;(\HOLBoundVar{st\ensuremath{\sb{1}}}\HOLSymConst{,}\HOLConst{Rval}\;\HOLBoundVar{res}))\\
\;\;\HOLTokenBar{}\;(\HOLBoundVar{st\ensuremath{\sb{1}}}\HOLSymConst{,}\HOLConst{Rerr}\;\HOLBoundVar{err})\;\HOLTokenImp{}\;(\HOLBoundVar{st\ensuremath{\sb{1}}}\HOLSymConst{,}\HOLConst{Rerr}\;\HOLBoundVar{err})
\end{holthmenv}
Evaluating an expression only changes the state by decrementing the clock when function calls are made, which we have proven as a theorem about the semantics.
Evaluation order for expressions follows Dafny's short-circuit semantics.
In the absence of short-circuiting, we choose to evaluate expressions from left to right, which feels more natural in the context of Dafny.

As a verification-oriented programming language, Dafny provides expressions that are only for verification.
These can be used to give hints to the automated verifier (with the \verb|assert| statement) or annotate methods with a pre- and postcondition (with keywords \verb|requires| and \verb|ensures| respectively).
In particular, the expressions in these constructs are required to be Dafny expressions that evaluate to boolean values.
The postcondition expression for many methods will often need to relate the state of the heaps before and after the execution of the method body.
Accordingly, Dafny has special support for \verb|old|-expressions, which allow referring to the state before the body was executed within the postcondition.

To illustrate this, consider the following method that swaps two elements in an array:
\begin{alltt}
method Swap(a: array<int>, i: int, j: int)
  requires 0 <= i < a.Length && 0 <= j < a.Length
  ensures a[i] == old(a[j]) && a[j] == old(a[i])
  modifies a
\{
  \graycomment{// <- old refers to the state here}
  var temp := a[i];
  a[i] := a[j];
  a[j] := temp;
  \graycomment{// <- ensures refers to state here}
\}
\end{alltt}
As \verb|ensures| refers to the state after executing the method body, the user must use \verb|old| to refer to the array before it was modified.

We define the semantics for \verb|old| as
\begin{holthmenv}
  \HOLConst{evaluate_exp}\;\HOLFreeVar{st}\;\HOLFreeVar{env}\;(\HOLConst{OldHeap}\;\HOLFreeVar{e})\;\HOLTokenDefEquality{}\\
\;\;\HOLKeyword{case}\;\HOLConst{evaluate_exp}\;(\HOLConst{use_old_heap}\;\HOLFreeVar{st})\;\HOLFreeVar{env}\;\HOLFreeVar{e}\;\HOLKeyword{of}\\
\;\;\;\;(\HOLBoundVar{st\ensuremath{\sb{1}}}\HOLSymConst{,}\HOLBoundVar{r})\;\HOLTokenImp{}\;(\HOLConst{unuse_old_heap}\;\HOLBoundVar{st\ensuremath{\sb{1}}}\;\HOLFreeVar{st}\HOLSymConst{,}\HOLBoundVar{r})
\end{holthmenv}
\begin{holthmenv}
  \HOLConst{use_old_heap}\;\HOLFreeVar{st}\;\HOLTokenDefEquality{}\;\HOLFreeVar{st}\;\HOLKeyword{with}\;\HOLFieldName{heap}\;:=\;\HOLFreeVar{st}.\HOLFieldName{heap_old} \\
  \HOLConst{unuse_old_heap}\;\HOLFreeVar{cur}\;\HOLFreeVar{prev}\;\HOLTokenDefEquality{}\;\HOLFreeVar{cur}\;\HOLKeyword{with}\;\HOLFieldName{heap}\;:=\;\HOLFreeVar{prev}.\HOLFieldName{heap}
\end{holthmenv}

In other words, a user-annotated expression wrapped within \verb|old| is evaluated in the heap state with respect to the old heap, which the semantics sets when it enters a callee.

We also track the old locals because the state does not separate track parameters, i.e., they are indistinguishable from other locals, and because Dafny allows shadowing parameters.
To illustrate this point, suppose that after swapping the elements in the array, the user writes \verb|var a := 0;| shadowing \verb|a| to be an integer.
If we were to take the \verb|ensures| as is, \verb|a| would mistakenly refer to the newly declared integer variable.
We avoid this by interpreting references to parameters in \verb|ensures| annotations to be implicitly wrapped in an \HOLConst{Old} expression, which, analogous to \HOLConst{OldHeap} above, evaluates an expression in the old heap and the old locals.

\subsection{Statements} \label{semantics:stmt}
As many Dafny programs are written in an imperative style, we also define semantics for statements, which may modify the state of the program.

\paragraph{Skip, Return, and Then}
We can easily define the semantics of \HOLConst{Skip}, \HOLConst{Return}, and \HOLConst{Then}:
\begin{holthmenv}
  \HOLConst{evaluate_stmt}\;\HOLFreeVar{st}\;\HOLFreeVar{env}\;\HOLConst{Skip}\;\HOLTokenDefEquality{}\;(\HOLFreeVar{st}\HOLSymConst{,}\HOLConst{Rcont}) \\
  \HOLConst{evaluate_stmt}\;\HOLFreeVar{st}\;\HOLFreeVar{env}\;\HOLConst{Return}\;\HOLTokenDefEquality{}\;(\HOLFreeVar{st}\HOLSymConst{,}\HOLConst{Rstop}\;\HOLConst{Sret}) \\
  \HOLConst{evaluate_stmt}\;\HOLFreeVar{st\ensuremath{\sb{0}}}\;\HOLFreeVar{env}\;(\HOLConst{Then}\;\HOLFreeVar{stmt\ensuremath{\sb{1}}}\;\HOLFreeVar{stmt\ensuremath{\sb{2}}})\;\HOLTokenDefEquality{}\\
\;\;\HOLKeyword{case}\;\HOLConst{evaluate_stmt}\;\HOLFreeVar{st\ensuremath{\sb{0}}}\;\HOLFreeVar{env}\;\HOLFreeVar{stmt\ensuremath{\sb{1}}}\;\HOLKeyword{of}\\
\;\;\;\;(\HOLBoundVar{st\ensuremath{\sb{1}}}\HOLSymConst{,}\HOLConst{Rcont})\;\HOLTokenImp{}\;\HOLConst{evaluate_stmt}\;\HOLBoundVar{st\ensuremath{\sb{1}}}\;\HOLFreeVar{env}\;\HOLFreeVar{stmt\ensuremath{\sb{2}}}\\
\;\;\HOLTokenBar{}\;(\HOLBoundVar{st\ensuremath{\sb{1}}}\HOLSymConst{,}\HOLConst{Rstop}\;\HOLBoundVar{stp})\;\HOLTokenImp{}\;(\HOLBoundVar{st\ensuremath{\sb{1}}}\HOLSymConst{,}\HOLConst{Rstop}\;\HOLBoundVar{stp}) \\
\end{holthmenv}
Recall that the clock is used to prove termination of the interpreter.
As the recursive call in the \HOLConst{Then} case takes a strictly smaller statement as an argument, we do not need to decrement the clock to prove termination.

\paragraph{Assign}
Dafny supports parallel assignment, meaning that it is possible to swap the value of two variables by writing, e.g.,
\begin{alltt}
                 x, y := y, x;
\end{alltt}
which we support by first evaluating the expressions on the right-hand side of an assignment, followed by assigning to the left-hand sides from left to right.

\paragraph{Array Allocation}
To allocate a (possibly nested) one-dimensional array in Dafny, we can write
\begin{alltt}
                a := new T[len];
\end{alltt}
where \verb|T| is the type of the array and \verb|len| its length.
Note that the allocation on the right-hand side of the assignment does not specify an initial value.
According to the Dafny reference, the initial array elements are arbitrary values of type \verb|T|.
This is in contrast to our semantics, which supplies a default value for the array elements.
For the types supported by our subset, defining a default value is straightforward.
The advantage of this tweak is that it avoids the non-deterministic choice of an arbitrary value in the semantics, which otherwise would have to be accounted for, e.g., by making the semantics non-deterministic via an oracle.

Formally, array allocation is defined as appending a heap value \HOLConst{HArr xs t} to the heap, where \HOLConst{xs} is a list of \HOLConst{len} copies of the default value, and returning a value \HOLConst{ArrV} which, as described in Section~\ref{semantics:value}, records the length, location on the heap, and the type of the array.
Note that Dafny does not support deallocation.

\paragraph{While}
We define the semantics for \HOLConst{While} as follows:\footnote{Note that we do not check that the loop invariants \texttt{invs} hold here. As we will see in Section~\ref{sec:vvcg}, they are considered in the wp-calculus instead. This suggests that our semantics for Dafny is actually a combination of the functional big-step semantics presented here and the sound wp-calculus.}
\begin{holthmenv}
  \HOLConst{evaluate_stmt}\;\HOLFreeVar{st\ensuremath{\sb{0}}}\;\HOLFreeVar{env}\\
\;\;(\HOLConst{While}\;\HOLFreeVar{guard}\;\HOLFreeVar{invs}\;\HOLFreeVar{decrs}\;\HOLFreeVar{mods}\;\HOLFreeVar{body})\;\HOLTokenDefEquality{}\\
\;\;\HOLKeyword{if}\;\HOLFreeVar{st\ensuremath{\sb{0}}}.\HOLFieldName{clock}\;\HOLSymConst{=}\;\HOLNumLit{0}\;\HOLKeyword{then}\;(\HOLFreeVar{st\ensuremath{\sb{0}}}\HOLSymConst{,}\HOLConst{Rstop}\;(\HOLConst{Serr}\;\HOLConst{Rtimeout}))\\
\;\;\HOLKeyword{else}\\
\;\;\;\;\HOLKeyword{case}\;\HOLConst{evaluate_exp}\;(\HOLConst{dec_clock}\;\HOLFreeVar{st\ensuremath{\sb{0}}})\;\HOLFreeVar{env}\;\HOLFreeVar{guard}\;\HOLKeyword{of}\\
\;\;\;\;\;\;(\HOLBoundVar{st\ensuremath{\sb{1}}}\HOLSymConst{,}\HOLConst{Rval}\;(\HOLConst{BoolV}\;\HOLConst{F}))\;\HOLTokenImp{}\;(\HOLBoundVar{st\ensuremath{\sb{1}}}\HOLSymConst{,}\HOLConst{Rcont})\\
\;\;\;\;\HOLTokenBar{}\;(\HOLBoundVar{st\ensuremath{\sb{1}}}\HOLSymConst{,}\HOLConst{Rval}\;(\HOLConst{BoolV}\;\HOLConst{T}))\;\HOLTokenImp{}\\
\;\;\;\;\;\;(\HOLKeyword{case}\;\HOLConst{evaluate_stmt}\;\HOLBoundVar{st\ensuremath{\sb{1}}}\;\HOLFreeVar{env}\;\HOLFreeVar{body}\;\HOLKeyword{of}\\
\;\;\;\;\;\;\;\;\;(\HOLBoundVar{st\ensuremath{\sb{2}}}\HOLSymConst{,}\HOLConst{Rcont})\;\HOLTokenImp{}\\
\;\;\;\;\;\;\;\;\;\;\;\HOLConst{evaluate_stmt}\;\HOLBoundVar{st\ensuremath{\sb{2}}}\;\HOLFreeVar{env}\\
\;\;\;\;\;\;\;\;\;\;\;\;\;(\HOLConst{While}\;\HOLFreeVar{guard}\;\HOLFreeVar{invs}\;\HOLFreeVar{decrs}\;\HOLFreeVar{mods}\;\HOLFreeVar{body})\\
\;\;\;\;\;\;\;\HOLTokenBar{}\;(\HOLBoundVar{st\ensuremath{\sb{2}}}\HOLSymConst{,}\HOLConst{Rstop}\;\HOLBoundVar{stp})\;\HOLTokenImp{}\;(\HOLBoundVar{st\ensuremath{\sb{2}}}\HOLSymConst{,}\HOLConst{Rstop}\;\HOLBoundVar{stp}))\\
\;\;\;\;\HOLTokenBar{}\;(\HOLBoundVar{st\ensuremath{\sb{1}}}\HOLSymConst{,}\HOLConst{Rval}\;\HOLBoundVar{guard\HOLTokenUnderscore{}v})\;\HOLTokenImp{}\;(\HOLBoundVar{st\ensuremath{\sb{1}}}\HOLSymConst{,}\HOLConst{Rstop}\;(\HOLConst{Serr}\;\HOLConst{Rfail}))\\
\;\;\;\;\HOLTokenBar{}\;(\HOLBoundVar{st\ensuremath{\sb{1}}}\HOLSymConst{,}\HOLConst{Rerr}\;\HOLBoundVar{err})\;\HOLTokenImp{}\;(\HOLBoundVar{st\ensuremath{\sb{1}}}\HOLSymConst{,}\HOLConst{Rstop}\;(\HOLConst{Serr}\;\HOLBoundVar{err}))
\end{holthmenv}
Note that, unlike \HOLConst{Then}, the recursive call takes the same while loop as an argument, meaning that the size of the statement being evaluated stays the same.
Thus, every time we enter the loop, we need to check and decrement the clock.

It is not strictly necessary to check and decrement the clock immediately on entry; instead, it is possible to check and decrement the clock right before the recursive call.
We decided to use the former, as it simplifies the correctness proof for the compiler by more closely matching the semantics of function calls, which is what loops compile to.

\paragraph{Method Call}
\begin{figure}
\centering
\begin{alltt}
method Find(a: array<int>, key: int)
  returns (i: int)
  ensures 0 <= i ==> i < a.Length && a[i] == key
  ensures i < 0 ==>
    forall k :: 0 <= k < a.Length ==> a[k] != key
\{
  i := 0;
  while i < a.Length
    invariant 0 <= i <= a.Length
    invariant
      forall k :: 0 <= k < i ==> a[k] != key
  \{
    if a[i] == key \{ return; \}
    i := i + 1;
  \}
  i := -1;
\}
\end{alltt}
\caption{Linear search in Dafny.\protect\footnotemark}
\label{code:dfy:search}
\end{figure}
\footnotetext{Adapted from \url{https://dafny.org/dafny/OnlineTutorial/guide}.}
We avoid showing the full semantics of a method call due to its length.
Instead, we use an example.
Suppose we want to call a method \verb|Find|, which implements linear search (Fig.~\ref{code:dfy:search}).
We write
\begin{alltt}
              idx := Find(a, key)
\end{alltt}
as the method call.
Note that the caller must assign all out-parameters.
The semantics of the method call are as follows:
\begin{enumerate}
\item{Look up ``Find'' in the \HOLConst{prog} field of the environment.}
\item{Evaluate the parameters \verb|a| and \verb|key| in the caller.}
\item{Check if the in- and out-parameters \verb|a|, \verb|key|, and \verb|i| are distinct.}
\item{Change \HOLConst{locals} to only contain the parameters \verb|a| and \verb|key| initialized with their values, and the uninitialized out-parameter \verb|i|.}
\item{Copy the changed \HOLConst{locals} and the current \HOLConst{heap} to \HOLConst{locals_old} and \HOLConst{heap_old}, respectively.}
\item{Check the clock, and time out if no more ticks remain.}
\item{Decrement the clock and evaluate the method body, expecting \HOLinline{\HOLConst{Rstop}\;\HOLConst{Sret}} as the return value.}
\item{Evaluate the out-parameter \verb|i| in the callee.}
\item{Restore \HOLConst{locals}, \HOLConst{locals_old}, and \HOLConst{heap_old} to the original values of the caller.}
\item{Assign the value of the out-parameter \verb|i| to \verb|idx|.}
\end{enumerate}
There are two noteworthy points regarding the evaluation of the method body in step 6.
First, evaluating the method body requires a recursive call to \HOLConst{evaluate_stmt}.
As the method body can be arbitrarily large, we must check and decrement the clock to be able to prove termination.
Second, we require the recursive call to return \HOLinline{\HOLConst{Rstop}\;\HOLConst{Sret}}; that is, the method body must end with a return statement.
Note that if a source program does not explicitly end with return (such as in Fig.~\ref{code:dfy:search}), the return statement is added automatically by our frontend.

\subsection{Programs}
Matching Dafny's notion of executable programs, we define the semantics of a program as first checking whether methods and functions have unique names, followed by calling ``Main'' from an initial state where \HOLConst{locals}, \HOLConst{heap}, and their old counterpart are initialized with the empty list.
\begin{holthmenv}
  \HOLConst{evaluate_program}\;\HOLFreeVar{ck}\;(\HOLConst{Program}\;\HOLFreeVar{members})\;\HOLTokenDefEquality{}\\
\;\;\HOLKeyword{if}\;\HOLSymConst{\HOLTokenNeg{}}\HOLConst{distinct}\;(\HOLConst{map}\;\HOLConst{member_name}\;\HOLFreeVar{members})\;\HOLKeyword{then}\\
\;\;\;\;(\HOLConst{init_state}\;\HOLFreeVar{ck}\HOLSymConst{,}\HOLConst{Rstop}\;(\HOLConst{Serr}\;\HOLConst{Rfail}))\\
\;\;\HOLKeyword{else}\\
\;\;\;\;\HOLConst{evaluate_stmt}\;(\HOLConst{init_state}\;\HOLFreeVar{ck})\\
\;\;\;\;\;\;(\HOLConst{mk_env}\;(\HOLConst{Program}\;\HOLFreeVar{members}))\\
\;\;\;\;\;\;(\HOLConst{MetCall}\;[]\;\HOLStringLitDG{Main}\;[])
\end{holthmenv}

\section{Verified Compiler} \label{sec:vcomp}
We now turn to describing our formally verified compiler from Dafny to CakeML, which, together with CakeML's formally verified compiler~\cite{cakeml-compiler}, yields a verified compilation pipeline for Dafny down to machine code.
Our overall Dafny compilation pipeline is as follows:
\begin{enumerate}
\item{Generate an S-expression from an extended version of Dafny's existing frontend.}
\item{Parse the S-expression into a Dafny AST in HOL4.}
\item{Remove \verb|assert| statements by replacing them with \HOLinline{\HOLConst{Skip}}.}
\item{Freshen the AST, which updates all variable names to be unique and start with ``v'', simplifying proofs.}
\item{Compile the freshened Dafny AST into a CakeML AST, discarding annotations.}
\item{Compile the CakeML AST to machine code in a verified manner using the CakeML compiler~\cite{cakeml-compiler}.}
\end{enumerate}
This pipeline is engineered to maintain compatibility with both Dafny tooling and the CakeML codebase and proofs.
We have successfully compiled and run machine code for 18 code examples, including Fibonacci, the 91 function, swapping two elements in an array, and binary search.

In the rest of this section, we focus on describing the Dafny-to-CakeML transformation, namely steps 3 and 4.
The compiler for these steps is defined as the function \HOLinline{\HOLConst{compile}}
\begin{holthmenv}
  \HOLConst{compile}\;\HOLFreeVar{dfy}\;\HOLTokenDefEquality{}\\
\;\;\HOLConst{from_program}\;(\HOLConst{freshen_program}\;(\HOLConst{remove_assert}\;\HOLFreeVar{dfy}))
\end{holthmenv}

The technical challenges in implementing and verifying \HOLinline{\HOLConst{compile}} include reconciling our choice of left-to-right evaluation order with CakeML's right-to-left order and handling the possibility of early returns from Dafny method calls.
We first explain our implementation, then present the compiler correctness statement and proof.

\subsection{Compiler Implementation}
Let us return to the linear search implementation shown in Figure~\ref{code:dfy:search} to describe our compiler.
This example showcases many key features of Dafny; in order to compile it, we need to map Dafny's variables, arrays, while loops, (early) returns, and methods to CakeML.
Note that neither \verb|ensures| nor \verb|invariant| clauses, nor their contents such as \verb|forall|, are compiled, as they are used only for verification purposes.

\paragraph{Variables}
We compile Dafny variables to CakeML references, which are mutable variables and thus semantically equivalent.
For example, \verb|i := i + 1| in Dafny is compiled to \verb|i := !i + 1|, where \verb|!| is the dereference operator in CakeML.
We postpone the discussion of variable declarations to the compilation of the method signature later in this section.

\paragraph{Arrays}
Dafny arrays are compiled to tuples that hold the array and its length for the reason mentioned in Footnote~\ref{fn:arrlen}.
For example, in Figure~\ref{code:dfy:search}, the array \verb|a| is represented by a corresponding CakeML value of type \verb|(int * int array)|.
Thus, \verb|a[i]| is compiled to \verb|Array.sub (snd (!a)) (!i)| and \verb|a.Length| is compiled to \verb|fst (!a)|.

\paragraph{While Loops}
We compile while loops as tail-recursive functions, matching the definition of \verb|while| in Standard ML~\cite{sml}.
The CakeML compiler performs tail-call optimization, i.e., tail-recursive functions do not require allocating a new stack frame with every call, meaning that there is no risk of our compilation of loops causing excessive stack allocation.
The loop in Figure~\ref{code:dfy:search} is compiled to:
\begin{alltt}
     let fun loop () =
       if !i < fst (!a) then (
         if Array.sub (snd (!a), !i) = !key
         then raise Return else ();
         i := !i + 1;
         loop ()
       ) else ()
     in loop (); ... end
\end{alltt}
Note that we call \verb|loop| at least once.
After entering the loop, we check whether the loop condition holds.
If it does, we execute the body and recursively tail-call the loop.
If the condition does not hold, the expression evaluates to \verb|()|.

Additionally, observe that the Dafny code returns immediately once the key is found.
In CakeML, we compile this as raising the \verb|Return| exception.
To ensure the existence of this exception, the compiler inserts
\begin{alltt}
               exception Return;
\end{alltt}
at the beginning of every program.
We postpone the discussion of how we handle the \verb|Return| exception from the perspective of the caller until later in the text.

The rest of the body after the loop is compiled as
\begin{alltt}
                  i := ~1;
                  raise Return
\end{alltt}
Recall that the final (non-early) return line is implicitly inserted by the Dafny frontend at the source level and thereby compiled the same way as all return calls to a CakeML \verb|Return| exception.
This insertion ensures that every method call explicitly returns, which is consistent with our semantics.

\paragraph{Methods} \label{sec:vcomp:method}
Before discussing the compilation of the rest of the method outside the body, it is useful to first discuss the compilation of method calls.
Suppose that \verb|idx| is a local variable, and \verb|a| and \verb|key| are bound to some array and integer, respectively.
Then, the method call
\begin{alltt}
              idx := Find(a, key)
\end{alltt}
is compiled to
\begin{alltt}
            let val t0 = dfy\_Find key a
            in idx := t0 end
\end{alltt}

Note how we assign to the left-hand side in a separate expression, prepend ``dfy\_'' to the method name, and reverse the order of arguments.
While separating the assignment is not necessary in this case, it is necessary when a method has multiple out-parameters, in which case, the method returns a tuple.
For example,
\begin{alltt}
             min, max := MinMax(a)
\end{alltt}
is compiled to
\begin{alltt}
       let val (t0, t1) = dfy\_Find key a
       in min := t0; max := t1 end
\end{alltt}
If a method does not have any out-parameters, its return value and the assignment become \verb|()|.

By prepending ``dfy\_'' to method names, combined with the fact that after the freshen pass all user-generated variable names start with ``v'', the compiler can generate additional names, such as \verb|t0| and \verb|t1| in the preceding examples, which (provably) do not change the semantics of the program.

We reverse the order of arguments to account for the difference in evaluation order between CakeML and Dafny.
Instead of reversing the order of arguments, we could have enforced a left-to-right evaluation order using let-bindings.
The disadvantage of this approach is that we need to generate as many internal variables as there are arguments and prove that they do not change the semantics.
While this is possible and we do similar proofs in other parts of the compiler, they are tedious.

Having a picture of how method calls are compiled, we can now discuss the compilation of the rest of the method outside the body.
In particular, the compiled method must return the values of the out-parameters at the end of the method.
As a method always explicitly returns, we can simply wrap the compiled body in an exception handler, which reads the out-parameters and returns their value:
\begin{alltt}
             <compiled body>
               handle Return => !i
\end{alltt}

The last remaining step is to compile the method signature:
\begin{alltt}
             fun dfy\_Find key a =
             let
               val key = ref key
               val a = ref a
               val i = ref 0
             in ... end
\end{alltt}

As the body refers to the parameters as variables, and we compile variables as references, we have to allocate references for \verb|key|, \verb|a|, and \verb|i|.
Note that \verb|key| and \verb|a| are initialized with the respective arguments, whereas the out-parameter \verb|i| (uninitialized in Dafny) is initialized to be 0.

In general, we compile variable declarations, including out-parameters, as references that are initialized with 0 regardless of their type.
This can produce CakeML code that does not pass CakeML's type checker but allows us to avoid an extra layer of boxing, e.g., by storing references to options.
However, in some cases, this compilation scheme still introduces unnecessary assignments.
In the future, we plan to implement a compiler pass that eliminates such redundant assignments where possible, which would allow further optimizations such as removing references in cases where a variable is only read.

Note that generating potentially type-incorrect CakeML is not necessarily a problem in terms of correctness for two reasons:
first, the correctness theorem of the CakeML compiler applies to all programs with well-defined semantics, regardless of their type correctness.
Second, as a valid Dafny program will always assign variables before reading them, the generated CakeML program will also have well-defined semantics.
We will formalize a more general version of this argument with the correctness of our compiler in Section~\ref{sec:comp:correct}.

\begin{figure}
\centering
\begin{alltt}
exception Return;

(* fst and snd are inlined in the compiler *)
fun fst (x, y) = x;
fun snd (x, y) = y;

fun dfy\_Find (key: int) (a: int * (int array)) =
let
  val key = ref key
  val a = ref a
  val i = ref 0
in
  (i := 0;
   let fun loop () =
     if !i < fst (!a) then (
       if Array.sub (snd (!a), !i) = !key
       then raise Return else ();
       i := !i + 1;
       loop ()
     ) else ()
   in
     loop ();
     i := ~1;
     raise Return
   end)
  handle Return => !i
end;
\end{alltt}
\caption{Result of compiling \texttt{Find} to CakeML}
\label{code:cml:find}
\end{figure}

Figure~\ref{code:cml:find} shows the final compilation result for \verb|Find|.

\subsection{Compiler Correctness} \label{sec:comp:correct}
Informally, our correctness theorem states that evaluating a Dafny program using Dafny's functional big-step semantics is equivalent to evaluating the compiled Dafny program using CakeML's functional big-step semantics, according to a notion of equivalence we define.

\subsubsection{Equivalence}
Recall from Section~\ref{sec:semantics} that the functional big-step semantics returns a state and a result.
Step-by-step, we will build up our notion of equivalence for values, results, states, and environment, which we will use for our correctness theorems.

\paragraph{Value Relation}
We define equivalence between Dafny (left) and CakeML (right) values using the inductive \HOLinline{\HOLConst{val_rel}}, whose definition is straightforward for integers, booleans, and strings.
For example, it is defined as follows for integers:
\begin{holthmenv}
  \HOLTokenTurnstile{}\;\HOLConst{val_rel}\;\HOLFreeVar{m}\;(\HOLConst{IntV}\;\HOLFreeVar{i})\;(\HOLConst{Litv}\;(\HOLConst{IntLit}\;\HOLFreeVar{i}))
\end{holthmenv}

Array values in Dafny are equivalent to a CakeML tuple containing the length of an array and its location in CakeML's store.
We require that the lengths match, converting between \verb|num| and \verb|int| as necessary, and that the locations are related through the map \verb|m|.
\begin{holthmenv}
  \HOLTokenTurnstile{}\;\HOLConst{lookup}\;\HOLFreeVar{m}\;\HOLFreeVar{loc}\;\HOLSymConst{=}\;\HOLConst{Some}\;\ensuremath{\HOLFreeVar{loc}\sp{\prime{}}}\;\HOLSymConst{\HOLTokenImp{}}\\
\;\;\;\;\;\HOLConst{val_rel}\;\HOLFreeVar{m}\;(\HOLConst{ArrV}\;\HOLFreeVar{len}\;\HOLFreeVar{loc}\;\HOLFreeVar{ty})\\
\;\;\;\;\;\;\;(\HOLConst{Tuplev}\;[\HOLConst{Litv}\;(\HOLConst{IntLit}\;(\HOLSymConst{\&}\HOLFreeVar{len}));\;\HOLConst{Loc}\;\HOLConst{T}\;\ensuremath{\HOLFreeVar{loc}\sp{\prime{}}}])
\end{holthmenv}

\paragraph{Result Relation}
The equivalence between Dafny and CakeML expression results is built on \HOLinline{\HOLConst{val_rel}}:
\begin{holthmenv}
  \HOLConst{exp_res_rel}\;\HOLFreeVar{m}\;(\HOLConst{Rval}\;\HOLFreeVar{dfy\HOLTokenUnderscore{}v})\;(\HOLConst{Rval}\;[\HOLFreeVar{cml\HOLTokenUnderscore{}v}])\;\HOLTokenDefEquality{}\\
\;\;\HOLConst{val_rel}\;\HOLFreeVar{m}\;\HOLFreeVar{dfy\HOLTokenUnderscore{}v}\;\HOLFreeVar{cml\HOLTokenUnderscore{}v}
\end{holthmenv}

For statements that execute normally in Dafny, we only require that the corresponding CakeML expression return some value.
In the case of return, we check that the correct \verb|Return| exception has been raised in CakeML:
\begin{holthmenv}
  \HOLConst{stmt_res_rel}\;\HOLConst{Rcont}\;(\HOLConst{Rval}\;\ensuremath{\HOLFreeVar{v}\sb{\mathrm{0}}})\;\HOLTokenDefEquality{}\;\HOLConst{T} \\
  \HOLConst{stmt_res_rel}\;(\HOLConst{Rstop}\;\HOLConst{Sret})\;(\HOLConst{Rerr}\;(\HOLConst{Rraise}\;\HOLFreeVar{val}))\;\HOLTokenDefEquality{}\\
\;\;\HOLConst{is_ret_exn}\;\HOLFreeVar{val}
\end{holthmenv}

\paragraph{State Relation}
We define the equivalence between a Dafny and CakeML state as follows:
\begin{holthmenv}
  \HOLConst{state_rel}\;\HOLFreeVar{m}\;\HOLFreeVar{l}\;\HOLFreeVar{s}\;\HOLFreeVar{t}\;\HOLFreeVar{cml\HOLTokenUnderscore{}env}\;\HOLTokenDefEquality{}\\
\;\;\HOLFreeVar{s}.\HOLFieldName{clock}\;\HOLSymConst{=}\;\HOLFreeVar{t}.\HOLFieldName{clock}\;\HOLSymConst{\HOLTokenConj{}}\;\HOLConst{array_rel}\;\HOLFreeVar{m}\;\HOLFreeVar{s}.\HOLFieldName{heap}\;\HOLFreeVar{t}.\HOLFieldName{refs}\;\HOLSymConst{\HOLTokenConj{}}\\
\;\;\HOLConst{locals_rel}\;\HOLFreeVar{m}\;\HOLFreeVar{l}\;\HOLFreeVar{s}.\HOLFieldName{locals}\;\HOLFreeVar{t}.\HOLFieldName{refs}\;\HOLFreeVar{cml\HOLTokenUnderscore{}env}.\HOLFieldName{v} \\
\end{holthmenv}

Here, \HOLinline{\HOLFreeVar{s}} and \HOLinline{\HOLFreeVar{t}} are the Dafny and CakeML states, respectively.
By requiring the clocks to be equal, we can prove that a Dafny program diverges if and only if the corresponding CakeML program diverges.

The \HOLConst{array_rel} relation requires that for every array on Dafny's heap, the mapping \HOLinline{\HOLFreeVar{m}} provides the location of the corresponding array on CakeML's store and additionally requires that their contents satisfy \HOLConst{val_rel}.

The \HOLConst{locals_rel} relation requires that for every local variable in Dafny, the mapping \HOLinline{\HOLFreeVar{l}} provides the corresponding reference in CakeML, which must be bound to the same name in the CakeML environment (\HOLinline{\HOLFreeVar{cml\HOLTokenUnderscore{}env}}).
If the local variable is initialized in Dafny, the values must be equivalent with respect to \HOLConst{val_rel}.

\paragraph{Environment Relation}
We define \HOLConst{env_rel} to specify both a notion of equivalence between Dafny's and CakeML's environments and additional well-formedness conditions on the environments.
First, we require the presence of basic functions and constructors such as \verb|True|, \verb|False|, and \verb|::| (cons) in CakeML.
Additionally, we require that the program in Dafny's environment be well-formed, and for each of its members, the corresponding function must exist in CakeML's environment.

\subsubsection{Compiler Correctness Statement}
We formalize the correctness statement described informally at the beginning of this section:
\begin{holthmenv}
  \HOLTokenTurnstile{}\;\HOLConst{evaluate_program}\;\HOLFreeVar{dfy\HOLTokenUnderscore{}ck}\;\HOLFreeVar{prog}\;\HOLSymConst{=}\;(\HOLFreeVar{s}\HOLSymConst{,}\HOLConst{Rcont})\;\HOLSymConst{\HOLTokenConj{}}\\
\;\;\;\HOLConst{compile}\;\HOLFreeVar{prog}\;\HOLSymConst{=}\;\HOLConst{inr}\;\HOLFreeVar{cml\HOLTokenUnderscore{}decs}\;\HOLSymConst{\HOLTokenImp{}}\\
\;\;\;\;\;\HOLSymConst{\HOLTokenExists{}}\HOLBoundVar{ck}\;\ensuremath{\HOLBoundVar{t}\sp{\prime{}}}\;\ensuremath{\HOLBoundVar{m}\sp{\prime{}}}\;\HOLBoundVar{r\HOLTokenUnderscore{}cml}.\\
\;\;\;\;\;\;\;\HOLConst{evaluate_decs}\;(\HOLConst{cml_init_state}\;\HOLFreeVar{ffi}\;(\HOLFreeVar{dfy\HOLTokenUnderscore{}ck}\;\HOLSymConst{\ensuremath{+}}\;\HOLBoundVar{ck}))\\
\;\;\;\;\;\;\;\;\;(\HOLConst{cml_init_env}\;\HOLFreeVar{ffi})\;\HOLFreeVar{cml\HOLTokenUnderscore{}decs}\;\HOLSymConst{=}\\
\;\;\;\;\;\;\;\;\;(\ensuremath{\HOLBoundVar{t}\sp{\prime{}}}\HOLSymConst{,}\HOLBoundVar{r\HOLTokenUnderscore{}cml})\;\HOLSymConst{\HOLTokenConj{}}\\
\;\;\;\;\;\;\;\HOLConst{state_rel}\;\ensuremath{\HOLBoundVar{m}\sp{\prime{}}}\;\HOLConst{empty}\;\HOLFreeVar{s}\;\ensuremath{\HOLBoundVar{t}\sp{\prime{}}}\;(\HOLConst{cml_init_env}\;\HOLFreeVar{ffi})\;\HOLSymConst{\HOLTokenConj{}}\\
\;\;\;\;\;\;\;\HOLConst{stmt_res_rel}\;\HOLConst{Rcont}\;\HOLBoundVar{r\HOLTokenUnderscore{}cml}
\end{holthmenv}

This is a forward simulation result, which suffices to show that the (deterministic) semantics of Dafny programs are preserved in the output CakeML programs.
Note that the correctness statement only talks about Dafny programs that terminate successfully (\HOLConst{Rcont}).
We can use the VCG to be sure that a given program successfully terminates, which is implied by the correctness of the VCG and the soundness of the weakest precondition calculus (Section~\ref{sec:vvcg:soundness}).
In Section~\ref{sec:together}, we will combine this fact with the compiler theorem in the context of the 91 function.
As indicated by the sum value, the Dafny-to-CakeML compiler can fail, for example, if the compiler is asked to compile a \verb|forall|-expression; compiler correctness only applies when it succeeded.

To understand why we existentially quantify over additional clock ticks \HOLinline{\HOLFreeVar{ck}} in the CakeML initial state, recall that \HOLConst{state_rel} requires Dafny's and CakeML's clocks to be the same.
However, in some cases it may be necessary to compile Dafny code into CakeML code that requires (finitely) more ticks in the semantics.
For example, a Dafny method with $n$ parameters is compiled to a curried function, meaning that a call to that method requires at least one tick in Dafny's semantics, whereas it requires at least $n$ ticks in CakeML's semantics.

\paragraph{Correctness of Statement Compilation}
A major component of the top-level correctness proof is the correctness of the compiler for statements.
Omitting some technical details, it is stated as
\begin{holthmenv}
  \HOLTokenTurnstile{}\;\HOLConst{evaluate_stmt}\;\HOLFreeVar{s}\;\HOLFreeVar{env\HOLTokenUnderscore{}dfy}\;\HOLFreeVar{stmt\HOLTokenUnderscore{}dfy}\;\HOLSymConst{=}\;(\ensuremath{\HOLFreeVar{s}\sp{\prime{}}}\HOLSymConst{,}\HOLFreeVar{r\HOLTokenUnderscore{}dfy})\;\HOLSymConst{\HOLTokenConj{}}\\
\;\;\;\HOLFreeVar{r\HOLTokenUnderscore{}dfy}\;\HOLSymConst{\HOLTokenNotEqual{}}\;\HOLConst{Rstop}\;(\HOLConst{Serr}\;\HOLConst{Rfail})\;\HOLSymConst{\HOLTokenConj{}}\\
\;\;\;\HOLConst{from_stmt}\;\HOLFreeVar{stmt\HOLTokenUnderscore{}dfy}\;\HOLFreeVar{lvl}\;\HOLSymConst{=}\;\HOLConst{inr}\;\HOLFreeVar{e\HOLTokenUnderscore{}cml}\;\HOLSymConst{\HOLTokenConj{}}\\
\;\;\;\HOLConst{state_rel}\;\HOLFreeVar{m}\;\HOLFreeVar{l}\;\HOLFreeVar{s}\;\HOLFreeVar{t}\;\HOLFreeVar{env\HOLTokenUnderscore{}cml}\;\HOLSymConst{\HOLTokenConj{}}\\
\;\;\;\HOLConst{env_rel}\;\HOLFreeVar{env\HOLTokenUnderscore{}dfy}\;\HOLFreeVar{env\HOLTokenUnderscore{}cml}\;\HOLSymConst{\HOLTokenConj{}}\;\HOLConst{is_fresh_stmt}\;\HOLFreeVar{stmt\HOLTokenUnderscore{}dfy}\;\HOLSymConst{\HOLTokenConj{}}\\
\;\;\;\HOLConst{no_shadow}\;(\HOLConst{set}\;(\HOLConst{map}\;\HOLConst{fst}\;\HOLFreeVar{s}.\HOLFieldName{locals}))\;\HOLFreeVar{stmt\HOLTokenUnderscore{}dfy}\;\HOLSymConst{\HOLTokenImp{}}\\
\;\;\;\;\;\HOLSymConst{\HOLTokenExists{}}\HOLBoundVar{ck}\;\ensuremath{\HOLBoundVar{t}\sp{\prime{}}}\;\ensuremath{\HOLBoundVar{m}\sp{\prime{}}}\;\HOLBoundVar{r\HOLTokenUnderscore{}cml}.\\
\;\;\;\;\;\;\;\HOLConst{evaluate}\;(\HOLFreeVar{t}\;\HOLKeyword{with}\;\HOLFieldName{clock}\;:=\;\HOLFreeVar{t}.\HOLFieldName{clock}\;\HOLSymConst{\ensuremath{+}}\;\HOLBoundVar{ck})\;\HOLFreeVar{env\HOLTokenUnderscore{}cml}\\
\;\;\;\;\;\;\;\;\;\;[\HOLFreeVar{e\HOLTokenUnderscore{}cml}]\;\HOLSymConst{=}\;(\ensuremath{\HOLBoundVar{t}\sp{\prime{}}}\HOLSymConst{,}\HOLBoundVar{r\HOLTokenUnderscore{}cml})\;\HOLSymConst{\HOLTokenConj{}}\\
\;\;\;\;\;\;\;\HOLConst{state_rel}\;\ensuremath{\HOLBoundVar{m}\sp{\prime{}}}\;\HOLFreeVar{l}\;\ensuremath{\HOLFreeVar{s}\sp{\prime{}}}\;\ensuremath{\HOLBoundVar{t}\sp{\prime{}}}\;\HOLFreeVar{env\HOLTokenUnderscore{}cml}\;\HOLSymConst{\HOLTokenConj{}}\;\HOLFreeVar{m}\;\HOLSymConst{\HOLTokenSubmap{}}\;\ensuremath{\HOLBoundVar{m}\sp{\prime{}}}\;\HOLSymConst{\HOLTokenConj{}}\\
\;\;\;\;\;\;\;\HOLConst{stmt_res_rel}\;\HOLFreeVar{r\HOLTokenUnderscore{}dfy}\;\HOLBoundVar{r\HOLTokenUnderscore{}cml}
\end{holthmenv}

Note the similarities to the top-level correctness statement.
Here, we prove that evaluating a Dafny statement and its compiled counterpart results in a state and result that are equivalent under their respective relations; we also permit the CakeML semantics to begin with more clock ticks.

The main difference is that the assumptions of this latter statement are more general.
In particular, we universally quantify over all states and environments that satisfy \HOLConst{state_rel} and \HOLConst{env_rel}, respectively.
Additionally, we assume the properties that are established by the freshen pass, namely that variable names are unique and begin with ``v'', which is captured by \HOLConst{is_fresh_stmt} and \HOLConst{no_shadow}.
This generality allows us to prove compiler correctness using the induction principle arising from the termination proof of the functional big-step semantics for Dafny.

\section{Verified VCG} \label{sec:vvcg}
This section presents our formally verified verification condition generator (VCG), which, given a Dafny program, produces a list of verification conditions as Dafny expressions.
Verifying the output expressions shows functional correctness of Dafny programs and, moreover, guarantees well-definedness as assumed by our compiler.
A key challenge is to design the VCG to support sound analysis for loop framing and termination, which, in particular, enables verification for subtle examples like the 91 function (Fig.~\ref{code:dfy:mccarthy}).

We approached the VCG formalization in two phases:
\begin{enumerate}
\item{Define a weakest precondition calculus for Dafny as an inductive relation in HOL4 and prove it sound with respect to the Dafny semantics.}
\item{Define a concrete VCG implementation function that checks the program and emits verification conditions as Dafny expressions; this step is proved correct with respect to the wp-calculus.}
\end{enumerate}

This two-phase methodology is also applied when extending the wp-calculus with new rules or refining existing ones.
By splitting the VCG into a calculus and a function, we can focus on clearly and concisely specifying the wp-rules in the calculus while leaving potential performance considerations to the implementation of the VCG function.
Mirroring our development process, we will first describe our calculus and its soundness, followed by the VCG and its correctness.

\subsection{Weakest Precondition Calculus}
We implement the wp-calculus as an 8-place inductive relation for Dafny statements:

\begin{alltt}
\centering\HOLConst{stmt_wp}\;\HOLFreeVar{m}\;\ensuremath{\HOLFreeVar{reqs}}\;\ensuremath{\HOLFreeVar{stmt}}\;\ensuremath{\HOLFreeVar{post}}\;\HOLFreeVar{ens}\;\HOLFreeVar{decs}\;\HOLFreeVar{mods}\;\HOLFreeVar{ls}
\end{alltt}

In order of appearance, the parameters are:
the set of available methods \HOLinline{\HOLFreeVar{m}},
the preconditions \HOLinline{\HOLFreeVar{reqs}},
a Dafny statement \HOLinline{\HOLFreeVar{stmt}},
the postconditions \HOLinline{\HOLFreeVar{post}} and \HOLinline{\HOLFreeVar{ens}},
a termination measure \HOLinline{\HOLFreeVar{decs}},
a list of locations that may be modified \HOLinline{\HOLFreeVar{mods}},
and a list of defined locals and their types \HOLinline{\HOLFreeVar{ls}}.
The parameters \HOLinline{\HOLFreeVar{reqs}}, \HOLinline{\HOLFreeVar{post}}, \HOLinline{\HOLFreeVar{ens}}, \HOLinline{\HOLFreeVar{decs}}, and \HOLinline{\HOLFreeVar{mods}} are lists of expressions, matching the type of user annotations in the input program.
The rules of our calculus are more easily understood if one considers the parameter \HOLinline{\HOLFreeVar{reqs}} as the output and all the other parameters as inputs, which, as we will see later, is mirrored by the implementation of the VCG.

We distinguish two kinds of postconditions, where the parameter \HOLinline{\HOLFreeVar{ens}} determines the conditions that must hold upon statement return (i.e., \HOLConst{Rstop}), while \HOLinline{\HOLFreeVar{post}} determines the conditions that must hold assuming we continue executing normally (i.e., \HOLConst{Rcont}).
This is necessary to support early-return semantics, and the difference is easily illustrated by the definition of \HOLConst{stmt_wp} for the \HOLConst{Return} and \HOLConst{Skip} cases, which requires their preconditions to be \HOLConst{ens} and \HOLConst{post}, respectively:
\begin{holthmenv}
  \HOLTokenTurnstile{}\;\HOLConst{stmt_wp}\;\HOLFreeVar{m}\;\HOLFreeVar{ens}\;\HOLConst{Return}\;\HOLFreeVar{post}\;\HOLFreeVar{ens}\;\HOLFreeVar{decs}\;\HOLFreeVar{mods}\;\HOLFreeVar{ls} \\
  \HOLTokenTurnstile{}\;\HOLConst{stmt_wp}\;\HOLFreeVar{m}\;\HOLFreeVar{post}\;\HOLConst{Skip}\;\HOLFreeVar{post}\;\HOLFreeVar{ens}\;\HOLFreeVar{decs}\;\HOLFreeVar{mods}\;\HOLFreeVar{ls}
\end{holthmenv}

We can also easily express the relation in the case of statement composition:
\begin{holthmenv}
  \HOLTokenTurnstile{}\;\HOLConst{stmt_wp}\;\HOLFreeVar{m}\;\ensuremath{\HOLFreeVar{pre}\sb{\mathrm{1}}}\;\ensuremath{\HOLFreeVar{s}\sb{\mathrm{1}}}\;\ensuremath{\HOLFreeVar{pre}\sb{\mathrm{2}}}\;\HOLFreeVar{ens}\;\HOLFreeVar{decs}\;\HOLFreeVar{mods}\;\HOLFreeVar{ls}\;\HOLSymConst{\HOLTokenConj{}}\\
\;\;\;\HOLConst{stmt_wp}\;\HOLFreeVar{m}\;\ensuremath{\HOLFreeVar{pre}\sb{\mathrm{2}}}\;\ensuremath{\HOLFreeVar{s}\sb{\mathrm{2}}}\;\HOLFreeVar{post}\;\HOLFreeVar{ens}\;\HOLFreeVar{decs}\;\HOLFreeVar{mods}\;\HOLFreeVar{ls}\;\HOLSymConst{\HOLTokenImp{}}\\
\;\;\;\;\;\HOLConst{stmt_wp}\;\HOLFreeVar{m}\;\ensuremath{\HOLFreeVar{pre}\sb{\mathrm{1}}}\;(\HOLConst{Then}\;\ensuremath{\HOLFreeVar{s}\sb{\mathrm{1}}}\;\ensuremath{\HOLFreeVar{s}\sb{\mathrm{2}}})\;\HOLFreeVar{post}\;\HOLFreeVar{ens}\;\HOLFreeVar{decs}\;\HOLFreeVar{mods}\;\HOLFreeVar{ls}
\end{holthmenv}

The verification condition lists are interpreted conjunctively, so to add a verification condition, we can simply prepend it to the precondition, as in the case of user assertions:
\begin{holthmenv}
  \HOLTokenTurnstile{}\;\HOLConst{stmt_wp}\;\HOLFreeVar{m}\;(\HOLFreeVar{e}\HOLSymConst{::}\HOLFreeVar{post})\;(\HOLConst{Assert}\;\HOLFreeVar{e})\;\HOLFreeVar{post}\;\HOLFreeVar{ens}\;\HOLFreeVar{decs}\;\HOLFreeVar{mods}\;\HOLFreeVar{ls}
\end{holthmenv}
Note that there is an implicit well-formedness check for verification conditions: since they are expressions, to prove their validity we must show that they evaluate to \HOLConst{True} in all states, which is impossible if an expression is not well-formed, as its evaluation would fail in some state.

\paragraph{Assign}
A classical wp-rule for assignment would perform variable substitution, which can be tricky to get right, in particular in the presence of parallel assignment.
Additionally, if variables are substituted for large expressions, the generated conditions can explode in size~\cite{compact-vc}.
We avoid both of these issues by using a let-expression instead, for which we have defined functional big-step semantics similar to the assignment statement.

To illustrate this, suppose we want to determine the weakest precondition for the statement
\begin{alltt}
             b := a + a + a;
             x, y := b + b + b, c;
\end{alltt}
and the postcondition \verb|x < y|.
In the case of variable substitution, the intermediate result stemming from \HOLConst{Then} is
\begin{alltt}
                 b + b + b < c
\end{alltt}
with the final result being
\begin{alltt}
     a + a + a + a + a + a + a + a + a < c
\end{alltt}

With our approach of using \verb|let|, the intermediate result is
\begin{alltt}
          let x = b + b + b, y = c in
            x < y
\end{alltt}
with the final result being
\begin{alltt}
          let b = a + a + a in
          let x = b + b + b, y = c in
            x < y
\end{alltt}

Formally, the complete rule for assignment is as follows:
\begin{holthmenv}
  \HOLTokenTurnstile{}\;\HOLConst{map}\;\HOLConst{fst}\;\HOLFreeVar{l}\;\HOLSymConst{=}\;\HOLConst{map}\;\HOLConst{VarLhs}\;\HOLFreeVar{ret\HOLTokenUnderscore{}names}\;\HOLSymConst{\HOLTokenConj{}}\\
\;\;\;\HOLConst{map}\;\HOLConst{snd}\;\HOLFreeVar{l}\;\HOLSymConst{=}\;\HOLConst{map}\;\HOLConst{ExpRhs}\;\HOLFreeVar{exps}\;\HOLSymConst{\HOLTokenConj{}}\;\HOLConst{distinct}\;\HOLFreeVar{ret\HOLTokenUnderscore{}names}\;\HOLSymConst{\HOLTokenConj{}}\\
\;\;\;\HOLConst{every}\;(\HOLTokenLambda{}\HOLBoundVar{v}.\;\HOLSymConst{\HOLTokenNeg{}}\HOLConst{mem}\;\HOLBoundVar{v}\;\HOLFreeVar{mods})\;\HOLFreeVar{ret\HOLTokenUnderscore{}names}\;\HOLSymConst{\HOLTokenConj{}}\\
\;\;\;\HOLConst{length}\;\HOLFreeVar{exps}\;\HOLSymConst{=}\;\HOLConst{length}\;\HOLFreeVar{ret\HOLTokenUnderscore{}names}\;\HOLSymConst{\HOLTokenConj{}}\\
\;\;\;\HOLConst{set}\;\HOLFreeVar{ret\HOLTokenUnderscore{}names}\;\HOLSymConst{\HOLTokenSubset{}}\;\HOLConst{set}\;(\HOLConst{map}\;\HOLConst{fst}\;\HOLFreeVar{ls})\;\HOLSymConst{\HOLTokenConj{}}\\
\;\;\;\HOLConst{get_types}\;\HOLFreeVar{ls}\;\HOLFreeVar{exps}\;\HOLSymConst{=}\;\HOLConst{inr}\;\HOLFreeVar{rhs\HOLTokenUnderscore{}tys}\;\HOLSymConst{\HOLTokenConj{}}\\
\;\;\;\HOLConst{get_types}\;\HOLFreeVar{ls}\;(\HOLConst{map}\;\HOLConst{Var}\;\HOLFreeVar{ret\HOLTokenUnderscore{}names})\;\HOLSymConst{=}\;\HOLConst{inr}\;\HOLFreeVar{lhs\HOLTokenUnderscore{}tys}\;\HOLSymConst{\HOLTokenConj{}}\\
\;\;\;\HOLFreeVar{lhs\HOLTokenUnderscore{}tys}\;\HOLSymConst{=}\;\HOLFreeVar{rhs\HOLTokenUnderscore{}tys}\;\HOLSymConst{\HOLTokenImp{}}\\
\;\;\;\;\;\HOLConst{stmt_wp}\;\HOLFreeVar{m}\;[\HOLConst{Let}\;(\HOLConst{ZIP}\;(\HOLFreeVar{ret\HOLTokenUnderscore{}names}\HOLSymConst{,}\HOLFreeVar{exps}))\;(\HOLConst{conj}\;\HOLFreeVar{post})]\\
\;\;\;\;\;\;\;(\HOLConst{Assign}\;\HOLFreeVar{l})\;\HOLFreeVar{post}\;\HOLFreeVar{ens}\;\HOLFreeVar{decs}\;\HOLFreeVar{mods}\;\HOLFreeVar{ls}
\end{holthmenv}

The precondition is now a \HOLConst{Let} expression that binds the names being assigned to their respective expressions in the scope of the postcondition.
The rule also performs several syntactic well-formedness and type correctness\footnote{While VCG and type checking are typically performed as separate passes, we have combined them into a single phase for this work.} checks on assignments, which is necessary for soundness; these checks are usually added/modified as we discovered their necessity when formalizing soundness of the calculus.
We omit further discussion of these checks for brevity.

\paragraph{Array Update}
We also support array updates such as
\begin{alltt}
                   a[idx] := e;
\end{alltt}

Similar to the assignment rule, the rule for array updates performs several syntactic well-formedness and type correctness checks.
In particular, it syntactically checks that \verb|a| is part of the \verb|modifies| clause.
As a consequence, only variables can be included in the \verb|modifies|, and not expressions such as \verb|a[idx]|; we aim to lift this restriction in the future.

Omitting some well-formedness checks, the weakest precondition for the example above is:
\begin{alltt}
        0 <= idx /\textbackslash idx < a.Length /\textbackslash
        SetPrev (ForallHeap [a]
          (a[Prev idx] = (Prev e) /\textbackslash
           forall i: int ::
             (i != (Prev idx) /\textbackslash
              0 <= i /\textbackslash i < a.Length ==>
                a[i] = PrevHeap (a[i]))
           ==> post))
\end{alltt}
The expressions \HOLConst{SetPrev}, \HOLConst{Prev}, \HOLConst{PrevHeap}, and \HOLConst{ForallHeap} are our extensions to provide sufficient expressivity for verification conditions.
The expression \HOLConst{ForallHeap} quantifies over all heaps where previously allocated locations outside \verb|a| remain unchanged, including heaps with newly allocated locations.
More specifically, the argument to \HOLConst{ForallHeap} is a list of locations that are havoced and thus must be a subset of what is provided in the \verb|modifies| clause.
By using \HOLConst{Prev} and \HOLConst{PrevHeap} within \HOLConst{SetPrev}, expressions are evaluated at the point of \HOLConst{SetPrev}, with \HOLinline{\HOLFreeVar{locals}} and \HOLinline{\HOLFreeVar{heap}} appropriately set.
The index must be wrapped in \HOLConst{Prev} to correctly handle cases such as \verb|a[a[idx]] := a[idx]|.
Together, the verification condition frames the unmodified state around the update.

\paragraph{Array Allocation}
To understand the wp-rule for array allocation, consider
\begin{alltt}
       a := new T[len]; <next statement>
\end{alltt}
Observe that the example above is a sequence of statements, with the first statement assigning the newly allocated array to \verb|a|.
This is because the wp-rule for array allocation is a special case of the wp-rule for statement composition:
subsequent statements may modify \verb|a|, so it must be added to the list of locations that the following statement may modify.
Note that requiring allocation to be followed by another statement does not limit expressivity: statements only occur in methods, which, as previously mentioned, must explicitly return using a \HOLinline{\HOLConst{Return}} statement, meaning that allocation never occurs on its own.

Again, omitting some well-formedness checks, the weakest precondition for the example above is:
\begin{alltt}
  0 <= len /\textbackslash
  SetPrev (ForallHeap [] (forall a: array<T> ::
    (a.Length = (Prev len) /\textbackslash
     let x = (Prev default_e) in
       (forall i: int ::
          (0 <= i /\textbackslash i < a.Length ==> a[i] = x))
     ==> wp_next_stmt)))
\end{alltt}
Note that it uses the constructs \HOLConst{Prev} and \HOLConst{ForallHeap}, which were introduced in the context of array updates.
However, in contrast to array updates, the list passed to \HOLConst{ForallHeap} is empty, meaning that, except for potentially newly allocated locations, the heap is unchanged.
Recall from Section~\ref{semantics:stmt} that we define arrays to be initialized with some default value.
This fact is expressed in the verification condition using the let-expression.
Finally, \HOLConst{wp_next_stmt} is the weakest precondition of the statement following the allocation.
It is the same as in the general rule for statement composition, except that \HOLinline{\HOLFreeVar{a}} has been added to its \HOLinline{\HOLFreeVar{mods}}.

\paragraph{While}
Fundamentally, our definition of the wp-rule for while loops is the same as the standard Hoare Logic rule with annotated invariants.
However, our definition must additionally account for termination and framing to ensure that the loop preserves properties of variables not modified by its body.
To illustrate these points, consider \verb|SumToN| (Fig.~\ref{code:dfy:sum}), which includes the \verb|decreases| clause automatically guessed by the Dafny frontend.
\begin{figure}
\centering
\begin{alltt}
method SumToN(n: int) returns (sum: int)
  requires n >= 0
  ensures sum == n * (n + 1) / 2
  ensures n >= 0
\{
  var i := 1;
  sum := 0;

  while i <= n
    invariant 1 <= i <= n + 1
    invariant sum == (i - 1) * i / 2
    decreases n - i
  \{
    sum := sum + i;
    i := i + 1;
  \}
\}
\end{alltt}
\caption{Summing to \texttt{n} in Dafny.}
\label{code:dfy:sum}
\end{figure}
Keeping it high-level, the weakest precondition for the while loop is:
\begin{alltt}
      invariant (holds on entry) /\textbackslash
      invariant is maintained /\textbackslash
      forall sum, i ::
        !(i <= n) /\textbackslash invariant ==> ensures
\end{alltt}

Note that in the final conjunct, where we prove the postcondition, we only quantify over \verb|sum| and \verb|i|, which we have determined by syntactically checking which locals are assigned to.
In particular, we do not quantify over \verb|n|, effectively framing the property \verb|n >= 0|.
Since the loop does not include a \verb|modifies| clause, the heap is unchanged.

To ensure that the loop terminates, we need to check whether the expression in \verb|decreases| actually decreases.
For this, we need to be able to refer to the value of \verb|decreases| at the beginning of an iteration, which we achieve by storing that value in fresh variables.

If we were to explicitly apply this transformation to the loop body in Figure~\ref{code:dfy:sum}, we would get
\begin{alltt}
             var prev_n_i := n - i;
             sum := sum + i;
             i := i + 1;
\end{alltt}
It should be noted that we do not explicitly apply this transformation.
Instead, the wp-calculus quantifies over a list of variables that are sufficiently unique.

Knowing this, we can now discuss where the \verb|decreases| check happens, namely as part of checking that the invariant is maintained
\begin{alltt}
        i <= n /\textbackslash invariant ==> body_wp
\end{alltt}
where \verb|body_wp| represents the weakest precondition of the loop body and the postcondition
\begin{alltt}
          invariant /\textbackslash
          n - i < prev_n_i /\textbackslash
          0 <= prev_n_i /\textbackslash 0 <= n - i
\end{alltt}

\paragraph{Method Call}
The weakest precondition for method calls must make sure that the precondition of the call is satisfied, recursive calls decrease the termination measure in \verb|decreases|, and that the postcondition of the method call implies the weakest precondition of the rest of the statements.
An (optional) \verb|modifies| clause indicates locations on the heap that may have changed.

For example, the weakest precondition of the first call in the 91 function (Fig.~\ref{code:dfy:mccarthy}) is
\begin{alltt}
  \graycomment{// precondition}
  let n = n + 11 in True /\textbackslash
  \graycomment{// decreases}
  (let n = n + 11 in 111 - n) < old(111 - n) /\textbackslash
  0 <= (let n = n + 11 in 111 - n) /\textbackslash
  0 <= old(111 - n) /\textbackslash
  \graycomment{// postcondition implies wp for the rest}
  forall tmp: int ::
    (let n = n + 11, r = tmp in
       r == if n <= 100 then 91 else n - 10)
     ==> ...
\end{alltt}

We can simplify the condition by noticing that, at the method call, \verb|n <= 100| holds, and that because the method does not update \verb|n|, we know that \verb|old(n)| is equal to \verb|n|.

Applying substitution and basic simplification rules to the last conjunct, we get:
\begin{alltt}
      forall tmp: int ::
        tmp == if n <= 89 then 91 else n + 1
        ==> ...
\end{alltt}
By considering the fact that \verb|n <= 100|, we know that \verb|tmp| is a number that is between 91 and 101.
Hence, it is less than \verb|111|, meaning that the value of \verb|111 - tmp| is non-negative.
Considering what we know about \verb|tmp| from the postcondition of the call, we also know that it is larger than \verb|n|.
Thus, the next recursive call will pass the \verb|decreases| check.
From the postcondition and the range of \verb|tmp|, we know that \verb|r| will be exactly 91, proving the postcondition of the method for \verb|n <= 100|.

\subsection{Weakest Precondition Soundness Statement} \label{sec:vvcg:soundness}
The soundness of our weakest precondition calculus consists of
a top-level soundness theorem for methods, which uses a soundness result for \HOLConst{stmt_wp}.

Informally, the soundness theorem for methods states that if for every method its precondition (\verb|requires|) implies the weakest precondition (\HOLConst{stmt_wp}) of its method body, then each method will adhere to its specification whenever started from a state satisfying its precondition.

This top-level method soundness theorem is proved by induction on a well-founded order in such a way that we get an inductive hypothesis stating that any method call using an input state that is smaller, according to the well-founded order, behaves according to its specification.
The order we use is a lexicographic order on the ``level'' of a method and the values of its \verb|decreases| clause.
The level of a method is determined using a topological sort on the call graph of a program.
Thus, if methods are (mutually) recursive, they have the same level and may only call each other if the values of the \verb|decreases| clause actually decrease at calls.
Otherwise, the callee has a lower level than the caller, and no \verb|decreases| clause is required.

The soundness proof of methods uses the following soundness theorem for the weakest precondition for statements.
One can read this soundness theorem as saying: if \HOLinline{\HOLFreeVar{reqs}} holds in the initial state \HOLinline{\HOLFreeVar{st}}, and all methods in \HOLinline{\HOLFreeVar{m}} can be found in the environment \HOLinline{\HOLFreeVar{env}}, then evaluating \HOLinline{\HOLFreeVar{stmt}} will terminate and result in a state where \HOLinline{\HOLFreeVar{post}} or \HOLinline{\HOLFreeVar{ens}} is satisfied, depending on whether the statement returns.
\begin{holthmenv}
\HOLTokenTurnstile{}\;\HOLConst{stmt_wp}\;\HOLFreeVar{m}\;\HOLFreeVar{reqs}\;\HOLFreeVar{stmt}\;\HOLFreeVar{post}\;\HOLFreeVar{ens}\;\HOLFreeVar{decs}\;\HOLFreeVar{ls}\;\HOLSymConst{\HOLTokenImp{}}\\
\;\;\;\HOLConst{conditions_hold}\;\HOLBoundVar{st}\;\HOLBoundVar{env}\;\HOLFreeVar{reqs}\;\HOLSymConst{\HOLTokenConj{}}\\
\;\;\;\HOLConst{compatible_env}\;\HOLBoundVar{env}\;\HOLFreeVar{m}\;\HOLSymConst{\HOLTokenConj{}}\;\ensuremath{\dots}\;\HOLSymConst{\HOLTokenImp{}}\\
\;\;\;\;\HOLSymConst{\HOLTokenExists{}}\ensuremath{\HOLBoundVar{st}\sp{\prime{}}}\;\HOLBoundVar{ret}.\\
\;\;\;\;\;\;\;\;\;\HOLConst{eval_stmt}\;\HOLBoundVar{st}\;\HOLBoundVar{env}\;\HOLFreeVar{stmt}\;\ensuremath{\HOLBoundVar{st}\sp{\prime{}}}\;\HOLBoundVar{ret}\;\HOLSymConst{\HOLTokenConj{}}\\
\;\;\;\;\;\;\;\;\;\HOLKeyword{case}\;\HOLBoundVar{ret}\;\HOLKeyword{of}\\
\;\;\;\;\;\;\;\;\;\;\;\HOLConst{Rcont}\;\HOLTokenImp{}\;\HOLConst{conditions_hold}\;\ensuremath{\HOLBoundVar{st}\sp{\prime{}}}\;\HOLBoundVar{env}\;\HOLFreeVar{post}\\
\;\;\;\;\;\;\;\;\;\HOLTokenBar{}\;\HOLConst{Rstop}\;\HOLConst{Sret}\;\HOLTokenImp{}\;\HOLConst{conditions_hold}\;\ensuremath{\HOLBoundVar{st}\sp{\prime{}}}\;\HOLBoundVar{env}\;\HOLFreeVar{ens}\\
\;\;\;\;\;\;\;\;\;\HOLTokenBar{}\;\HOLConst{Rstop}\;(\HOLConst{Serr}\;\ensuremath{\HOLBoundVar{v}\sb{\mathrm{3}}})\;\HOLTokenImp{}\;\HOLConst{F}\;\HOLSymConst{\HOLTokenConj{}}\\
\;\;\;\;\;\;\;\;\;\ensuremath{\dots}
\end{holthmenv}
Note that this is a total correctness result---the fact that evaluating \HOLinline{\HOLFreeVar{stmt}} will terminate is hidden in the \HOLinline{\HOLConst{eval_stmt}} relation (more details below), which requires there to exist clocks such that evaluation does not time out.

The soundness theorem above has some of its assumptions omitted (\ldots) because the assumptions include the lengthy inductive hypothesis regarding method calls: specifically, that we can assume that every method that we might call adheres to its specification if the call is made in a state where the well-founded order used in the method soundness proof has decreased.
The fact that the measure decreases is something we get to know from the verification conditions we generate for each call site.

To prove the soundness of \HOLConst{stmt_wp}, we proceed by induction over the definition of \HOLConst{stmt_wp}, showing that each wp-rule is locally sound.
For its proof, it is more convenient to work with a relational-style semantics, as opposed to the functional big-step style used in the compiler proof.

To bridge the relational and functional styles, we defined the relations \HOLConst{eval_exp} and \HOLConst{eval_stmt} (shown below), which effectively quotient out the functional big-step clocks in \HOLConst{evaluate_exp} and \HOLConst{evaluate_stmt}, respectively, by existential quantification.
\begin{holthmenv}
  \HOLConst{eval_exp}\;\HOLFreeVar{st}\;\HOLFreeVar{env}\;\HOLFreeVar{e}\;\HOLFreeVar{v}\;\HOLTokenDefEquality{}\\
\;\;\HOLSymConst{\HOLTokenExists{}}\ensuremath{\HOLBoundVar{ck}\sb{\mathrm{1}}}\;\ensuremath{\HOLBoundVar{ck}\sb{\mathrm{2}}}.\\
\;\;\;\;\HOLConst{evaluate_exp}\;(\HOLFreeVar{st}\;\HOLKeyword{with}\;\HOLFieldName{clock}\;:=\;\ensuremath{\HOLBoundVar{ck}\sb{\mathrm{1}}})\;\HOLFreeVar{env}\;\HOLFreeVar{e}\;\HOLSymConst{=}\\
\;\;\;\;\;\;(\HOLFreeVar{st}\;\HOLKeyword{with}\;\HOLFieldName{clock}\;:=\;\ensuremath{\HOLBoundVar{ck}\sb{\mathrm{2}}}\HOLSymConst{,}\HOLConst{Rval}\;\HOLFreeVar{v})\\
  \HOLConst{eval_stmt}\;\HOLFreeVar{st}\;\HOLFreeVar{env}\;\HOLFreeVar{body}\;\ensuremath{\HOLFreeVar{st}\sp{\prime{}}}\;\HOLFreeVar{ret}\;\HOLTokenDefEquality{}\\
\;\;\HOLSymConst{\HOLTokenExists{}}\ensuremath{\HOLBoundVar{ck}\sb{\mathrm{1}}}\;\ensuremath{\HOLBoundVar{ck}\sb{\mathrm{2}}}.\\
\;\;\;\;\HOLConst{evaluate_stmt}\;(\HOLFreeVar{st}\;\HOLKeyword{with}\;\HOLFieldName{clock}\;:=\;\ensuremath{\HOLBoundVar{ck}\sb{\mathrm{1}}})\;\HOLFreeVar{env}\;\HOLFreeVar{body}\;\HOLSymConst{=}\\
\;\;\;\;\;\;(\ensuremath{\HOLFreeVar{st}\sp{\prime{}}}\;\HOLKeyword{with}\;\HOLFieldName{clock}\;:=\;\ensuremath{\HOLBoundVar{ck}\sb{\mathrm{2}}}\HOLSymConst{,}\HOLFreeVar{ret})\;\HOLSymConst{\HOLTokenConj{}}\\
\;\;\;\;\HOLFreeVar{ret}\;\HOLSymConst{\HOLTokenNotEqual{}}\;\HOLConst{Rstop}\;(\HOLConst{Serr}\;\HOLConst{Rtimeout})
\end{holthmenv}
This allows us to make further relational big-step-style definitions and also to prove lemmas about these relations.
For example, given an expression \HOLinline{\HOLFreeVar{e}}, \HOLConst{eval_true} asserts that there exist clocks such that \HOLinline{\HOLFreeVar{e}} evaluates to true;
in fact, \mbox{\HOLConst{conditions_hold}} (used in the statement of the soundness theorem) is also formally defined with respect to this relation.
\begin{holthmenv}
  \HOLConst{eval_true}\;\HOLFreeVar{st}\;\HOLFreeVar{env}\;\HOLFreeVar{e}\;\HOLTokenDefEquality{}\;\HOLConst{eval_exp}\;\HOLFreeVar{st}\;\HOLFreeVar{env}\;\HOLFreeVar{e}\;(\HOLConst{BoolV}\;\HOLConst{T})\\
  \HOLConst{conditions_hold}\;\HOLFreeVar{st}\;\HOLFreeVar{env}\;\HOLTokenDefEquality{}\;\HOLConst{every}\;(\HOLConst{eval_true}\;\HOLFreeVar{st}\;\HOLFreeVar{env})
\end{holthmenv}
We can state the relational big-step-style semantics for the ``then'' branch of an if-statement as
\begin{holthmenv}
  \HOLTokenTurnstile{}\;\HOLConst{eval_true}\;\HOLFreeVar{st}\;\HOLFreeVar{env}\;\HOLFreeVar{grd}\;\HOLSymConst{\HOLTokenConj{}}\;\HOLConst{eval_stmt}\;\HOLFreeVar{st}\;\HOLFreeVar{env}\;\HOLFreeVar{thn}\;\HOLFreeVar{st\ensuremath{\sb{1}}}\;\HOLFreeVar{ret}\;\HOLSymConst{\HOLTokenImp{}}\\
\;\;\;\;\;\HOLConst{eval_stmt}\;\HOLFreeVar{st}\;\HOLFreeVar{env}\;(\HOLConst{If}\;\HOLFreeVar{grd}\;\HOLFreeVar{thn}\;\HOLFreeVar{els})\;\HOLFreeVar{st\ensuremath{\sb{1}}}\;\HOLFreeVar{ret}
\end{holthmenv}
Similar relational big-step style lemmas about the definitions \HOLConst{eval_exp} and \HOLConst{eval_stmt} are used throughout the soundness proof.

\subsection{Verified VCG Implementation}
The final step in our development of a verified VCG for Dafny is to define a function that generates verification conditions.
The implementation follows naturally from the definition of the wp-calculus and is written in a result monad.
For example, several clauses of the generator are shown below, corresponding to the wp-calculus snippets defined earlier.
\begin{holthmenv}
  \HOLConst{stmt_vcg}\;\HOLFreeVar{m}\;\HOLConst{Return}\;\HOLFreeVar{post}\;\HOLFreeVar{ens}\;\HOLFreeVar{decs}\;\HOLFreeVar{mods}\;\HOLFreeVar{ls}\;\HOLTokenDefEquality{}\;\HOLConst{inr}\;\HOLFreeVar{ens} \\
  \HOLConst{stmt_vcg}\;\HOLFreeVar{m}\;\HOLConst{Skip}\;\HOLFreeVar{post}\;\HOLFreeVar{ens}\;\HOLFreeVar{decs}\;\HOLFreeVar{mods}\;\HOLFreeVar{ls}\;\HOLTokenDefEquality{}\;\HOLConst{inr}\;\HOLFreeVar{post} \\
  \HOLConst{stmt_vcg}\;\HOLFreeVar{m}\;(\HOLConst{Then}\;\HOLFreeVar{s\ensuremath{\sb{1}}}\;\HOLFreeVar{s\ensuremath{\sb{2}}})\;\HOLFreeVar{post}\;\HOLFreeVar{ens}\;\HOLFreeVar{decs}\;\HOLFreeVar{mods}\;\HOLFreeVar{ls}\;\HOLTokenDefEquality{}\\
\;\;\HOLKeyword{case}\;\HOLConst{dest_ArrayAlloc}\;\HOLFreeVar{s\ensuremath{\sb{1}}}\;\HOLKeyword{of}\\
\;\;\;\;\HOLConst{None}\;\HOLTokenImp{}\\
\;\;\;\;\;\;\HOLKeyword{do}\\
\;\;\;\;\;\;\;\;\ensuremath{\HOLBoundVar{pre}\sp{\prime{}}}\;\HOLTokenLeftmap{}\;\HOLConst{stmt_vcg}\;\HOLFreeVar{m}\;\HOLFreeVar{s\ensuremath{\sb{2}}}\;\HOLFreeVar{post}\;\HOLFreeVar{ens}\;\HOLFreeVar{decs}\;\HOLFreeVar{mods}\;\HOLFreeVar{ls};\\
\;\;\;\;\;\;\;\;\HOLConst{stmt_vcg}\;\HOLFreeVar{m}\;\HOLFreeVar{s\ensuremath{\sb{1}}}\;\ensuremath{\HOLBoundVar{pre}\sp{\prime{}}}\;\HOLFreeVar{ens}\;\HOLFreeVar{decs}\;\HOLFreeVar{mods}\;\HOLFreeVar{ls}\\
\;\;\;\;\;\;\HOLKeyword{od}\\
\;\;\HOLTokenBar{}\;\HOLConst{Some}\;\ensuremath{\dots}\;\HOLTokenImp{}\ensuremath{\dots}\\
\HOLConst{stmt_vcg}\;\HOLFreeVar{m}\;(\HOLConst{Assert}\;\HOLFreeVar{e})\;\HOLFreeVar{post}\;\HOLFreeVar{ens}\;\HOLFreeVar{decs}\;\HOLFreeVar{mods}\;\HOLFreeVar{ls}\;\HOLTokenDefEquality{}\\
\;\;\HOLConst{inr}\;(\HOLFreeVar{e}\HOLSymConst{::}\HOLFreeVar{post})\\
    \HOLConst{stmt_vcg}\;\HOLFreeVar{m}\;(\HOLConst{Assign}\;\HOLFreeVar{ass})\;\HOLFreeVar{post}\;\HOLFreeVar{ens}\;\HOLFreeVar{decs}\;\HOLFreeVar{mods}\;\HOLFreeVar{ls}\;\HOLTokenDefEquality{}\\
\;\;\HOLKeyword{let}\;(\HOLBoundVar{lhss}\HOLSymConst{,}\HOLBoundVar{rhss})\;=\;\HOLConst{UNZIP}\;\HOLFreeVar{ass}\\
\;\;\HOLKeyword{in}\\
\;\;\;\;\HOLKeyword{do}\\
\;\;\;\;\;\;\HOLBoundVar{vars}\;\HOLTokenLeftmap{}\;\HOLConst{result_mmap}\;\HOLConst{dest_VarLhs}\;\HOLBoundVar{lhss};\\
\;\;\;\;\;\;\HOLBoundVar{es}\;\HOLTokenLeftmap{}\;\HOLConst{result_mmap}\;\HOLConst{dest_ExpRhs}\;\HOLBoundVar{rhss};\\
\;\;\;\;\;\;\HOLConst{assert}\;(\HOLConst{distinct}\;\HOLBoundVar{vars})\\
\;\;\;\;\;\;\;\;\HOLStringLitDG{stmt_vcg:Assign: variables not distinct};\\
\;\;\;\;\;\;\HOLConst{assert}\;(\HOLConst{list_disjoint}\;\HOLBoundVar{vars}\;\HOLFreeVar{mods})\\
\;\;\;\;\;\;\;\;\HOLStringLitDG{stmt_vcg:Assign: assigning to mods};\\
\;\;\;\;\;\;\ensuremath{\dots}\\
\;\;\;\;\;\;\HOLConst{inr}\;[\HOLConst{Let}\;(\HOLConst{ZIP}\;(\HOLBoundVar{vars}\HOLSymConst{,}\HOLBoundVar{es}))\;(\HOLConst{conj}\;\HOLFreeVar{post})]\\
\;\;\;\;\HOLKeyword{od}
\end{holthmenv}
Note that in the case of \HOLConst{Then}, we split on whether the first statement allocates an array, matching our description of the wp-rule for array allocation as a special case of statement composition.

We proved the correctness of \HOLConst{stmt_vcg} by showing that it only produces verification conditions that can be derived using the wp-calculus:
\begin{holthmenv}
  \HOLTokenTurnstile{}\;\HOLConst{stmt_vcg}\;\HOLFreeVar{m}\;\HOLFreeVar{stmt}\;\HOLFreeVar{post}\;\HOLFreeVar{ens}\;\HOLFreeVar{decs}\;\HOLFreeVar{mods}\;\HOLFreeVar{ls}\;\HOLSymConst{=}\;\HOLConst{inr}\;\HOLFreeVar{res}\;\HOLSymConst{\HOLTokenImp{}}\\
\;\;\;\;\;\HOLConst{stmt_wp}\;(\HOLConst{set}\;\HOLFreeVar{m})\;\HOLFreeVar{res}\;\HOLFreeVar{stmt}\;\HOLFreeVar{post}\;\HOLFreeVar{ens}\;\HOLFreeVar{decs}\;\HOLFreeVar{mods}\;\HOLFreeVar{ls}
\end{holthmenv}

While we have focused on the VCG for statements in this section, our mechanization also includes a formalization of the VCG for methods.
In particular, the generated verification condition for methods requires that the \verb|requires| clause of a method imply the weakest precondition of the body.

\section{Putting It All Together} \label{sec:together}
To demonstrate that our formally verified tools work in unison, we have used them to compile and verify the 91 function in an end-to-end manner.
As a brief reminder, the complete workflow is as follows.

\begin{enumerate}
\item We use an (untrusted) Dafny frontend to produce an S-expression for the 91 function.
\item The S-expression is parsed in HOL4, and we use HOL4's in-logic evaluation to generate verification conditions from \HOLConst{stmt_vcg}.
\item Since the verification condition is a Dafny expression, we prove its validity by showing that the expression evaluates to \HOLConst{True} in all states.
We have mechanized this proof, which involves expanding the semantics and using HOL4's decision procedure for integers.
\item Finally, we combine the validity of the verification condition, the soundness of the wp-calculus, the correctness of the VCG, and the correctness of the compiler to prove that the generated CakeML function satisfies the specification of the 91 function:
\end{enumerate}
\begin{holthmenv}
  \HOLTokenTurnstile{}\;\HOLConst{compile_member}\;\HOLConst{mccarthy}\;\HOLSymConst{=}\;\HOLConst{inr}\;\HOLFreeVar{mccarthy\HOLTokenUnderscore{}cml}\;\HOLSymConst{\HOLTokenConj{}}\\
\;\;\;\HOLConst{clos_env_ok}\;\HOLFreeVar{clos\HOLTokenUnderscore{}env}\;\HOLSymConst{\HOLTokenImp{}}\\
\;\;\;\;\;\HOLConst{AppReturns}\;(\HOLConst{INT}\;\HOLFreeVar{n})\\
\;\;\;\;\;\;\;(\HOLConst{Recclosure}\;\HOLFreeVar{clos\HOLTokenUnderscore{}env}\;[\HOLFreeVar{mccarthy\HOLTokenUnderscore{}cml}]\;\HOLStringLit{dfy_M})\\
\;\;\;\;\;\;\;(\HOLConst{INT}\;(\HOLKeyword{if}\;\HOLFreeVar{n}\;\HOLSymConst{\HOLTokenLeq{}}\;\HOLNumLit{100}\;\HOLKeyword{then}\;\HOLNumLit{91}\;\HOLKeyword{else}\;\HOLFreeVar{n}\;\HOLSymConst{\ensuremath{-}}\;\HOLNumLit{10}))
\end{holthmenv}

The assumption \HOLinline{\HOLConst{clos_env_ok}} ensures that basic functions and constructors are available to the method, while the predicate \HOLinline{\HOLConst{AppReturns}}~\cite{chargueraudphd} is defined in the context of the CakeML translator~\cite{proof-producing-jfp} and best understood as a Hoare triple.
Here, it says that applying the (recursive) closure of the CakeML 91 function to an input CakeML integer $n$ returns a CakeML integer that is either $91$ if $n \leq 100$ or $n - 10$ otherwise.
Notice that this result is verified entirely for CakeML program semantics and does not involve the Dafny toolchain(s).

We can apply both the compiler and the VCG to other methods, including the previously presented methods \verb|Swap|, \verb|Find|, and \verb|SumToN|.
Out of these methods, we have also proved the validity of the verification condition for \verb|Swap|.
We have not attempted to prove the verification conditions of the other methods, as the manual approach does not scale well.
Our aim is to improve this in future work by adding a path to SMT, allowing the use of SMT solvers to automatically prove the validity of verification conditions.

Similarly, we hope to minimize the proof effort needed to carry the functional correctness guarantees of verified Dafny methods to their CakeML counterparts, as we have manually done for the 91 function.

\section{Related Work}
We start with an overview of related work at the intersection of formalization and verification-aware programming platforms like Dafny, F*, and Why/Why3.
This is followed by a discussion of the broader literature on verified compilation and verified verification condition generation/checking.

\subsection{Verified Verification-Aware Programming}
\subsubsection*{Dafny}
To the best of our knowledge, verification condition generation and compilation for Dafny has not been done in a foundational way before.
Nezamabadi and Myreen~\cite{bakingfordafny} implemented a Dafny-to-CakeML compiler for a subset of Dafny that is larger than the one we present here, but only proved correctness for the compilation of some binary expressions.
There have also been efforts to implement Dafny's compilers in Dafny~\cite{compiler-bootstrap}, with the initial target being a purely functional subset of Dafny.

Dafny targets the intermediate verification language Boogie~\cite{boogie, boogie2} in order to generate verification conditions; there has been work by Parthasarathy et al.~\cite{formal-boogie} on validating Boogie's outputs using Isabelle/HOL.

Gladshtein et al.~\cite{lean-dafny} present a framework for foundational, multi-modal program verifiers, which they instantiate, among other things, to reason about Dafny-style programs.
In contrast to the deep embedding presented in this work, they use a monadic shallow embedding in Lean.

\subsubsection*{F*}
F*~\cite{fstar} is a proof-oriented programming language that, like Dafny, supports automated reasoning about programs using SMT solvers.
However, while Dafny mostly follows an imperative style, F* encourages a higher-order functional programming with effects style and has a dependent type system.
Strub et al.~\cite{fstar-self-certify} have implemented a type checker for F* in F* that returns type derivations.
In particular, it can return the type derivation for typechecking itself, which in turn can be checked in Rocq.

F* is also a popular target for embedding domain-specific programming languages:
Low*~\cite{karamel} is a low-level programming language that is shallowly embedded into F* and has a paper formalization of extracting it to Clight~\cite{clight}, which can be compiled by the CompCert~\cite{compcert} compiler, a verified C compiler that has been formalized in Rocq.
Steel~\cite{steel} is a concurrent programming language that is shallowly embedded into F*.
It is based on the concurrent separation logic SteelCore~\cite{steelcore} and has verified verification condition generation mechanized in F*.
There has also been work on embedding a subset of the x64 assembly language in F*~\cite{x64-fstar}, which includes an embedding of Vale~\cite{vale}, an automated verifier for high-performance assembly code that builds on top of existing verification languages like Dafny and F*, and a verified and efficient VCG.
To our knowledge, these embeddings have not been verified in a foundational proof assistant.

\subsubsection*{Why3 and Why}
Why3~\cite{why3} is a platform for deductive program verification.
It provides an ML dialect called WhyML for which it can generate verification conditions.
Unlike Dafny, it does not automatically discharge the generated verification conditions.
Instead, users can choose between automated and interactive provers.
To generate simpler verification conditions, it statically controls aliasing using its type system~\cite{why3-typesystem}.
Why is a verification condition generator for a ``WHILE'' language~\cite{why}.
Similar to Why3, Why can output its verification condition to automated and interactive provers.
It has inspired work by Herms et al.~\cite{certified-multi-prover}, where they have implemented a VCG that can produce verification conditions for multiple provers and proven it sound in Rocq.

\subsection{Verified Compilation and Verified Verification Condition Generation}

\subsubsection*{Combining Compilation and Verification}
The Verified Software Toolchain (VST)~\cite{vst} project combines CompCert with static analysis tools for invariant generation and verified invariant checking for an end-to-end verified toolchain for C programs.
It has been used, for example, to verify the OpenSSL implementation of SHA-256~\cite{openssl-sha256} and HMAC with SHA-256~\cite{openssl-hmac}.

CakeML provides a verified implementation of Characteristic Formulae for ML~\cite{cfml, cfcml-live, cfcml-io, cfcml}, which allows users to state specifications in the style of separation logic and prove them correct using tactics within HOL4.

seL4~\cite{sel4} is a verified microkernel written in C that has been formalized in Isabelle/HOL. Its compilation down to machine code has been proven correct using a translation validation approach~\cite{sel4-compile}.
More specifically, it makes use of a proof-producing decompilation of the binary~\cite{decompile,decompile-improved}, a formalization of the machine architecture~\cite{armv7}, and SMT solvers to show that the binary is a refinement of its C semantics. It has also made extensive use of the VCG provided by Simpl to prove refinement between its executable, monad specification, and high-performance C implementation~\cite{mind-the-gap}.

\subsubsection*{Verification Condition Generation}
Over the years, there have been multiple works on verification condition generation that were built as tactics in interactive theorem provers such as HOL or Isabelle:
Gordon~\cite{vcg-hol-gordon} presented a shallow embedding of a simple imperative language, for which he mechanically derived a Hoare logic and developed tactics to generate verification conditions.

Agerholm~\cite{vcg-hol-sten} presented a shallow embedding of an imperative language with non-determinism and loops, for which he formalized weakest precondition semantics and implemented a verification condition generator using tactics.

Homeier and Martin~\cite{vcg-hol} presented a deep embedding of a standard while loop language including expressions with side effects, for which they proved sound axiomatic semantics and used the axiomatic semantics to define a verified verification condition generator.

Huisman and Jacobs~\cite{huisman2000java} formalized the semantics of an imperative language including loops and abrupt termination (e.g., return and continue) in higher-order logic, provided an extension of Hoare logic with abrupt termination, and used it in the context of Java with the proof tool PVS.
Its approach is reminiscent of our combination of functional big-step semantics and a weakest precondition calculus.

Schirmer~\cite{simpl} presented Simpl, a model for sequential imperative programming languages, including a proven sound and complete Hoare logic for both partial and total correctness and a verification condition generator embedded as an Isabelle tactic.

There has also been work on verifying optimized verification condition generation:
Vogels et al.~\cite{vvcgen-efficient} used Rocq to formalize an efficient VC generation algorithm~\cite{leino-wp, compact-vc} that avoids an exponential blowup.
Grégoire and Sacchini~\cite{vc-static-analysis} formalized in Rocq the use of static analysis to simplify generated verification conditions for Java bytecode.

\subsubsection*{Verified End-to-End Compilers}

PureCake~\cite{purecake} is a verified compiler for a Haskell-like language, which, similar to our Dafny compiler, is built on top of CakeML.
Pancake~\cite{pancake} is a systems programming language with a verified compiler, which reuses the lower parts of the CakeML backend.

CertiCoq~\cite{certicoq} compiles Gallina, the specification language of Rocq, to C, which can be compiled using CompCert to form an end-to-end verified compiler.
Note that while CertiCoq's components have been verified, proving a composed end-to-end correctness theorem is still ongoing work~\cite{certicoq-ongoing}.

V\'elus~\cite{compile-lustre} is a verified compiler for synchronous dataflow languages like Lustre~\cite{lustre} and is built on top of CompCert.

In the context of the CompCert project, the translation validation approach has been employed for an SSA-based middle-end~\cite{ssa}, instruction scheduling~\cite{instruction-scheduling}, and software pipelining~\cite{software-pipelining}.

\section{Conclusion}
We have presented a verified compiler and verified verification condition generator based on functional big-step semantics for an imperative subset of Dafny.
Our subset includes mutually recursive method calls, while loops, and arrays, which is expressive enough to support interesting examples such as McCarthy's 91 function and array-based programs that are used when teaching Dafny.
Together, our formalization shows that it is possible to obtain foundational correctness guarantees across the entire toolchain for verification-aware programming languages.

\subsection*{Future Work}
Beyond the contributions described in this paper, we believe that our work provides a scaffolding that can be expanded in different directions: feature support, automation, and integration with the CakeML ecosystem.

By improving feature support and automation, we can bring our implementation closer to a drop-in replacement for the current Dafny toolchain.
Initial steps in this regard could be to support Dafny's compiler litmus test suite and to implement and verify a connection to SMT, allowing the use of automatic SMT solvers.
Supporting Dafny's compiler litmus test suite will most likely involve building support for Dafny's object-oriented features (traits and classes) as well as functional features (first-class function values and algebraic data types).
The latter features have straightforward compilation targets in CakeML, and ideas to compile traits and classes have been very briefly sketched in previous work~\cite{bakingfordafny}.

Furthermore, a deeper integration of our work with the CakeML ecosystem could allow CakeML code to call Dafny code, and vice versa.
Thus, it may enable the input and output of Dafny programs to be placed on solid foundations.
More generally, a program could simultaneously make use of Dafny's verification automation, CakeML's HOL4 frontend~\cite{proof-producing-jfp, proof-producing-monadic}, and Hoare-style reasoning using separation logic~\cite{cfcml, cfcml-io, cfcml-live}, without compromising on trustworthiness.

\begin{acks}
  We thank Clément Pit-Claudel for early discussions of this work, Fabio Madge for answering our questions about Dafny, Rustan Leino for discussions about Dafny at LICS 2025, and the anonymous reviewers for their comments.
  This work was supported by an Amazon Research Award and NTU Singapore's Global Connect Fellowship.
  The second author was partially funded by Swedish Research Council grant 2021-05165.
\end{acks}

\bibliographystyle{ACM-Reference-Format}
\bibliography{paper}

\end{document}